\documentclass[a4paper,11pt]{elsarticle}
\usepackage[utf8]{inputenc}

\usepackage{lineno}
\journal{submitted to arXiv}

\usepackage[margin=2.0cm]{geometry}
\usepackage{booktabs}
\usepackage{amsmath,amssymb,amsfonts}
\usepackage{hyperref}
\usepackage{bm}
\hypersetup{
    colorlinks=true,
    linkcolor=blue,
    filecolor=magenta,      
    urlcolor=cyan!80!blue,
    citecolor = green!80!black, 
    }

\usepackage{siunitx}
\usepackage{xcolor}
\usepackage{algorithm}
\usepackage{algpseudocode}
\usepackage{mathrsfs}
\usepackage{enumitem}
\usepackage{xcolor}
\usepackage{multicol}
\usepackage{multirow}
\usepackage{caption}
\usepackage{subcaption}

\usepackage{xfrac} %\sfrac{1}{2}

\usepackage[most]{tcolorbox}

\tcbset{highlight style/.style={
  enhanced,
  colback=yellow!30,
  colframe=yellow!30,
  boxrule=0pt,
  sharp corners,
  borderline west={1mm}{0pt}{yellow},
  left=1pt,
  right=1pt,
  top=1pt,
  bottom=1pt,
}}

\let\oldequation\equation
\let\oldendequation\endequation

\renewenvironment{equation}
  {\linenomathNonumbers\oldequation}
  {\oldendequation\endlinenomath}

\begin{document}

\begin{frontmatter}

\title{Numerical modelling of a partially loaded intermodal container freight train passing through a tunnel}

\author[BH]{Zhen Liu\corref{cor1}} 
\ead{15581666545@163.com}
\author[BH]{David Soper}
\author[BH]{Hassan Hemida} 
\author[BH]{Boyang Chen\corref{cor1}} 
\ead{bc2015@ic.ac.uk}
\cortext[cor1]{Corresponding author}

\address[BH]{School of Engineering, University of Birmingham, Birmingham, B15 2TT, UK}

\begin{abstract}
The bluff nature of a freight train locomotive, coupled with large gaps created between different wagon formations and loaded goods, influence the overall pressure wave pattern generated as the train passes through a tunnel. Typically, 1D models are used to predict the patterns and properties of tunnel pressure wave formations. However, accurate modelling of regions of separation at the head of the blunted containers and at unloaded gap sections is essential for precise predictions of pressure magnitudes. This has traditionally been difficult to capture with 1D models. Furthermore, achieving this accuracy through 3D computational methods demands exceptional mesh quality, significant computational resources, and the careful selection of numerical models. This paper evaluates various numerical models to capture these complexities within regions of flow separation. Findings have supported the development of a new 1D programme to calculate the pressure wave generated by a freight locomotive entering a tunnel, and is here further extended to consider the discontinuities of the train body created by intermodal container loading patterns, by implementing new mesh system and boundary conditions into the 1D programme. A parameterisation study for different loading configurations is also presented to improve the overall programme adaptability, and the relationship between predetermined parameters and gap length is investigated. \textcolor{black}{We validate the effectiveness of the improved 1D model through comprehensive Large Eddy Simulation (LES) results and conduct an extensive parameterisation study to enhance its applicability across various loading configurations. Consequently, } this research bridges the gap in freight train tunnel aerodynamics, offering a versatile 1D numerical tool for accurate pressure wave prediction.
\end{abstract}

\begin{keyword} 
Freight train \sep Computational Fluid Dynamics \sep One-dimensional numerical method \sep Train/tunnel aerodynamics \sep Large Eddy Simulation 
\end{keyword}

\end{frontmatter}

\section{Introduction}
As a train passes through a railway tunnel, it generates a series of pressure waves that travel at the speed of sound within the tunnel. These pressure waves not only exert aerodynamic loads on the train body and tunnel infrastructure, but also adversely affect the comfort of passengers on the train and other trains within the tunnel, as well as individuals within the tunnel. Numerous studies have been conducted to consider the aerodynamics of trains in tunnels \cite{howe2003influence,liu2010aerodynamic,iliadis2020numerical,liu2020investigation,cross2015validated}. However, the majority have focused on passenger trains due to the speed at which these vehicles travel and the aerodynamic issues that occur at these higher speeds \cite{yoon2001prediction}. Although freight trains travel at slower speeds in relation to passenger rolling stock, the shape of a freight train typically looks very different to that of a passenger train \cite{soper2016aerodynamics}, with predominantly highly bluff features and discontinuities in train body shape, created by intermodal container loading patterns or the coupling region between different wagons in the train consist. These create large regions of separated flow, which can lead to increased peak magnitudes for pressure waves.

 As a freight train passes through a tunnel, it generates a unique characteristic set of pressure wave patterns that differ from passenger trains. Indeed, the term freight train covers a wide cross-section of rolling stock. For the purpose of this research, the term freight from therein relates to a locomotive hauling series of flatbed intermodal wagons loaded with International Organisiation for Standardisation (ISO) intermodal shipping containers. 
 
 The influence of the blunted nose on tunnel pressure waves has been previously numerically and experimentally studied by \citet{liu2023study,iliadis2019experimental}. The separation bubble generated at the blunted train head increases the effective blockage area. This effect has also been investigated by conducting a parameterisation study and implementing the results numerically into a modified 1D code \citep{liu2023study,liu2022effect}. Moreover, the front face of the container after large gaps between other containers/freight wagons has also been observed to generate subsequent pressure waves, due to changes in the effective blockage area from large scale regions of separated flow and thus influence the pressure wave pattern \cite{iliadis2019experimental}. In traditional 1D pressure wave models, the influence of the train body on pressure variation is calculated using frictional coefficient, train length and cross-sectional area \cite{woods1981generalised,vardy1999estimation,william1993theoretical}. This approach does not appropriately capture the discontinuities of a freight train in the pressure wave formation. Therefore, although traditional 1D methodologies have the prominent advantage of being fast and accurate when predicting pressure variation inside tunnels for passenger trains, in many cases these are not suitable for highly bluff and discontinuous freight train formations, especially for partially loaded intermodal freight trains. 

Aerodynamically there is a notable distinction between container-type freight trains and passenger trains through the formation of wake vortices due to boundary layer separation at the rear of the loaded containers. Current studies on container type intermodal freight trains have mainly focused on the influence of loading configuration on pressure and velocity distributions in the open air, as well as force coefficients \citep{soper2014experimental,flynn2014detached,kocon2021critical,li2017flow,giappino2018high,watkins1992aerodynamic}. \citet{maleki2019flow} used the embedded-LES(ELES) approach to simulate the flow around a scale freight train loaded with different configurations of double-stacked containers. Increasing the front or base gap sizes around the test wagon increased the train drag, with the front gap having a stronger effect. The near-wake consisted of symmetric and asymmetric vortices. Furthermore, when considering the gap size between loaded containers, it was shown that a front gap size of approximately 1.77 times the wagon width allowed for flow across to the downstream wagon. However, a front gap size of 3.23 times the wagon width caused wake closure and impinged the shear layers, resulting in higher drag. Increasing the front gap size while keeping the base gap constant contracted the recirculation zone and increased the curvature of the shear layers \citep{maleki2019flow,maleki2017assessment}. \citet{osth2014study} used Large Eddy Simulation (LES) to analyse the flow around a generic container freight wagon model. Two cases were numerically studied: a single wagon and a wagon submerged in a train set, by using a periodic boundary condition. The study observed oscillations in the separation bubble and upper shear layer, synchronised by vorticity waves. Counter-rotating vortices in the wagon gaps reduced drag in the train set, inhibited lateral oscillations, and caused dominant oscillations in the side force signal. \textcolor{black}{Recent studies have expanded understanding of container-specific aerodynamics: \citet{huo2022effect} analysed the impact of different vehicle formations on aerodynamic forces; \citet{giappino2018high} conducted wind tunnel tests on container wagons to obtain aerodynamic coefficients; \citet{kedare2015computational} used CFD to evaluate the flow around empty freight wagons; and \cite{lai2008optimizing} proposed operational strategies to minimise aerodynamic drag in intermodal train design. Furthermore, \citet{iliadis2019aerodynamics}and \citet{iliadis2019experimental} combined moving model experiments and CFD to investigate flow separation and wave propagation around container freight trains entering tunnels. These studies collectively highlight the need to account for the unique aerodynamic characteristics of partially loaded container configurations.}

The flow characteristics observed at the gap of partially loaded containers closely resemble those of a line of bluff bodies in close-proximity. Research efforts have been dedicated to studying the wake interference phenomenon (buffeting) arising from the arrangement of two or more bluff bodies with varying sizes and gap lengths. Notably, \citet{havel2001buffeting} conducted experimental investigations on surface-mounted cubes with varying the gap length, under a Reynolds number of 22,000. Three regimes are summarised and briefly described as a bi-stable regime at a gap length of approximately 1.5W \textcolor{black}{(characteristic width of the bluff body)}, a lock-in regime occurring at a gap length of 1.5W–2.3W and a quasi-isolated regime characterised by a large gap separation. Similar flow regimes are identified and analysed for various upstream and downstream dimensions of different bluff bodies \cite{hangan1999buffeting,havel2001buffeting,zdravkovich1977review}. While these studies offer valuable reference data for sharp-edged three-dimensional objects like freight wagons, most of these studies are conducted at significantly lower Reynolds number than those encountered in train flows. With the development of computational ability, simulations have been conducted to investigate the flow field around platoon of vehicles running in close-proximity. \textcolor{black}{While studies focusing specifically on partially loaded container freight trains remain limited in the literature, similar aerodynamic features—such as gap-induced flow separations and wake interactions—have been observed in platoons of bluff bodies such as lorries. Therefore, these physically analogous cases can help inform the expected flow behaviour in freight train configurations with large container gaps. 
\citet{uystepruyst2013flow} simulated the flow around a four-vehicle platoon using LES method when one of the platoon members undergoes in-line oscillations, where 1:20 reduced-scale cubes are used to represent the vehicles. \citet{he2019detached} and \citet{chen2021Dynamic} further simulate the flow around a platoon of bluff lorries and trains using DES numerical model. The drag coefficient, pressure distribution, slipstream and vortex characteristics are analysed, identifying a highly turbulent flow on the top and sides of the trailing lorries.}

 It is clear that to successfully numerically predict the formation of pressure waves induced by partially loaded freight trains passing through tunnels will require the accurate simulation of the regions of separated flow. Consequently, meticulous considerations regarding mesh precision, computational resources, and the selection of an appropriate numerical model will become paramount. \textcolor{black}{The primary innovation of this research lies in developing a novel 1D numerical model tailored specifically for accurately capturing the complex aerodynamic interactions arising from discontinuities in partially loaded intermodal container freight trains entering tunnels. Unlike traditional 1D models, which inadequately address these discontinuities and resulting flow separations, our model incorporates a newly devised mesh system and enhanced boundary conditions. This enables accurate simulation of the pressure waves influenced by separated flows around the blunt edges of containers and gaps between them. }  Firstly, Section.~\ref{chap:Numerical_Simulation} details the 1D and 3D numerical models used in this paper. A new mesh system and governing equations are introduced in the 1D code to address challenges caused by discontinuities on the train body due to partial loading configurations. \textcolor{black}{In Section~\ref{chap:reference_case}, the modified 1D model is validated through comparison with LES results. To enhance the applicability of the model, a parameterisation study involving three different loading configurations is presented. Section~\ref{chap: parameterisationStudy_loadingconfiguration} analyses and summarises the characteristic flow regions and streamline patterns around partially loaded freight trains. Based on the numerical observations, Section~\ref{chap:parameterisation_equation} formulates parameterisation equations that describe the separation region and its relationship with loading configurations. Finally, Section~\ref{chap:validation} validates the proposed parameterisation and the 1D model through an independent simulation.} The predictive capabilities of this 1D programme are shown to be greatly enhanced and are now capable of appropriately considering the challenges posed by partially loaded container freight trains. \textcolor{black}{This research significantly advances the predictive capabilities for freight train tunnel aerodynamics, providing an efficient yet accurate computational tool beneficial for preliminary design and operational planning in railway engineering.}

\section{Numerical Simulation}
\label{chap:Numerical_Simulation}
\textcolor{black}{The one-dimensional numerical methods used in this study are based on the modelling framework developed in our previous work~\cite{liu2023study}, which examined the impact of separation bubbles around freight train noses on tunnel pressure waves. The same numerical setups and mesh resolution adopted for the 3D simulations of a freight locomotive entering a tunnel—previously presented and validated in~\citep{liu2023study}—are also employed here. For completeness, the essential components of the methodology are briefly summarised in this section. In this work, we extend the framework to capture additional aerodynamic discontinuities introduced by partially loaded container configurations. Specifically, new boundary condition formulations and refined mesh strategies are proposed and validated to model the effects of these structural discontinuities along the partially loaded train body.}

\subsection{Geometry}
\begin{figure}[htbp]
    \centering   
    \begin{subfigure}[t]{0.72\textwidth}
    \centering
    \includegraphics[width=\textwidth]{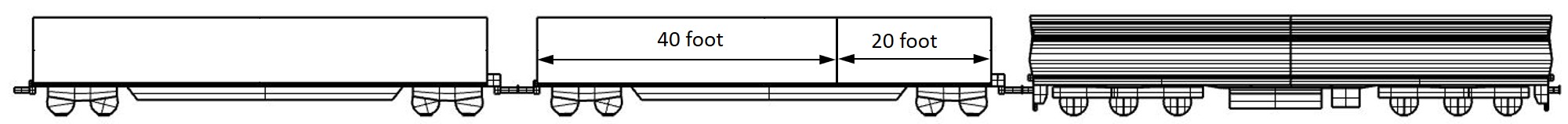}
    \caption*{\text{\small (a)}}
    \end{subfigure}
    \begin{subfigure}[t]{0.72\textwidth}
    \centering
    \includegraphics[width=\textwidth]{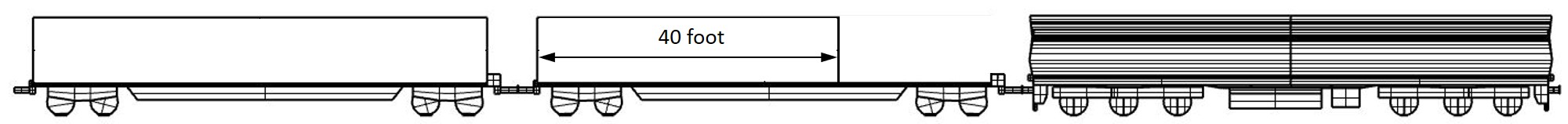}
    \caption*{\text{\small (b)}}
    \end{subfigure}
    \begin{subfigure}[t]{0.72\textwidth}
    \centering
    \includegraphics[width=\textwidth]{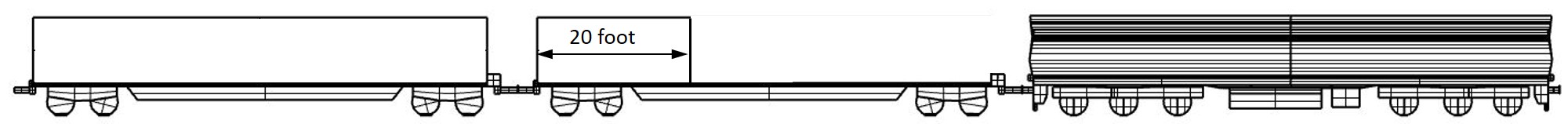}
    \caption*{\text{\small (c)}}
    \end{subfigure}
    \begin{subfigure}[t]{0.72\textwidth}
    \centering
    \includegraphics[width=\textwidth]{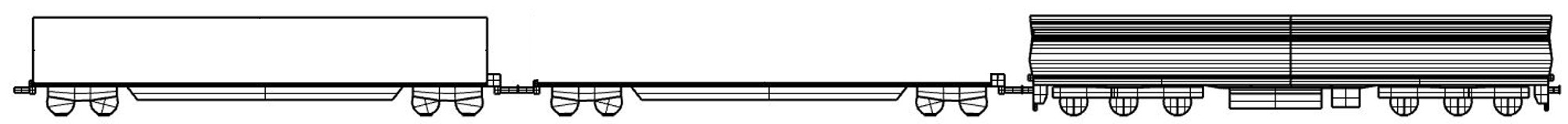}
    \caption*{\text{\small (d)}}
    \end{subfigure}
    \caption{Different loading configurations adopted for the parameterisation study (a) fully loaded train (b) 20 foot gap (c) 40 foot gap (d) 60 foot gap (1 carriage empty).}
    \label{fig:Geometry_parameterisation_loading}
\end{figure}

This study utilises a 1/25 scale model of a Class 66 freight locomotive connected to simplified FEA-B wagons loaded with ISO containers. The locomotive has a scale height of $H = 0.157~m$ and width of $W = 0.107~m$. For the parameterisation study of container loading configurations on the intermodal flatbed wagons, four loading configurations are used in this investigation. The flatbed wagon chosen for this study is the FEA-B type, with a \SI{18.2}{m} (60 foot) loading deck, as used in previous aerodynamic studies \citep{soper2014experimental,iliadis2019experimental}. According to standard International Shipping Organisation (ISO) designs, two types of container are designed with a size of \SI{12.192}{m} $\times$ \SI{2.438}{m} $\times$ \SI{2.590}{m} (40 foot) and \SI{6.096}{m} $\times$ \SI{2.438}{m} $\times$ \SI{2.590}{m} (20 foot). Thus by adjusting the position and the number of containers, four loading configurations are obtained to study the influence of the gap size, ranging from 0 m to 18.2 m (0 to 60 foot), on the flow parameters in the gaps (Figure~\ref{fig:Geometry_parameterisation_loading}). More specifically, the first working condition used both a 40 foot and a 20 foot containers, known as the fully loaded case. The second and third working conditions used one 40 foot and one 20 foot container for the first carriage of the train, which is empty for the fourth case, as shown in \textcolor{black}{Figure~\ref{fig:Geometry_parameterisation_loading}}. 

\subsection{1D model}

\subsubsection{Numerical method}
A 1D programme has been developed to calculate the pressure and velocity of the aerodynamic flow for a train passing through a tunnel using an unsteady, compressible, non-isentropic flow model \cite{woods1981generalised,vardy1976use,mei2013generalized}. This model accounts for heat transfer and friction effects between the air and the train/tunnel walls, with the air inside the tunnel modelled as an ideal gas. The continuity, momentum and energy equations are given as Equation.~\ref{continuity_1D}, Equation.~\ref{momentum_1D} and Equation.~\ref{energy_1D}.

\begin{equation}
\frac{\partial \rho}{\partial t}+u \frac{\partial \rho}{\partial x}+\rho \frac{\partial u}{\partial x}=0 
\label{continuity_1D}
\end{equation}

\begin{equation}
    \frac{\partial u}{\partial t}+u \frac{\partial u}{\partial x}+\frac{1}{\rho} \frac{\partial p}{\partial x}+\bm{g}=0
    \label{momentum_1D}  
\end{equation}

\begin{equation}
\left(\frac{\partial p}{\partial t}+u \frac{\partial p}{\partial x}\right)-a^2\left(\frac{\partial \rho}{\partial t}+u \frac{\partial \rho}{\partial x}\right)=(\bm{q}-\bm{w}+u \bm{g}) \rho(\kappa-1)
\label{energy_1D}
\end{equation}
\textcolor{black}{where $\rho$, $p$ and $u$ are the density, pressure and velocity of the local air, $\kappa$ is the ratio of specific heats}, $a$ is the local speed of sound, $\bm{g}$ is the friction term, $\bm{q}$ is the heat-transfer term and $\bm{w}$ denotes to the work-transfer term. The Method of Characteristics (MOC), a well-known mathematical technique, is employed to solve the quasi-linear hyperbolic partial differential equations, \textcolor{black}{by introducing three Riemann variables $\lambda$, $\beta$, and $A_A$. These are defined as follows:
\begin{equation}
\lambda = \frac{c}{c_R} + \frac{\kappa - 1}{2} \frac{u}{c_R}, \quad
\beta = \frac{c}{c_R} - \frac{\kappa - 1}{2} \frac{u}{c_R}, \quad
A_A = \frac{a_A}{c_R}
\end{equation}
Here, $c$ is the local speed of sound, $u$ is the local flow velocity, and $c_R$ is the reference speed of sound. $a_A$ is defined as the speed of sound when the airflow undergoes isentropic expansion to the reference pressure. It is a function of the entropy of the gas $s$, calculated from Equation~\ref{eq:aA},
\begin{equation}
a_A=c_{R}e^{(\frac{s-s_R}{2C_p})}
\label{eq:aA}
\end{equation}
All variables are non-dimensionalised by their respective reference values, making the Riemann variables themselves non-dimensional. $\lambda$ and $\beta$ represent the amplitudes of the forward and backward travelling wave characteristics, and are used as the main variables in the compatibility equations along the characteristic lines. $A_A$ represents the entropy characteristic. These serve as intermediate variables in the MOC method to calculate physical quantities such as pressure, velocity, and density.
}

\begin{figure}[H]
    \centering   
    \includegraphics[width=10cm]{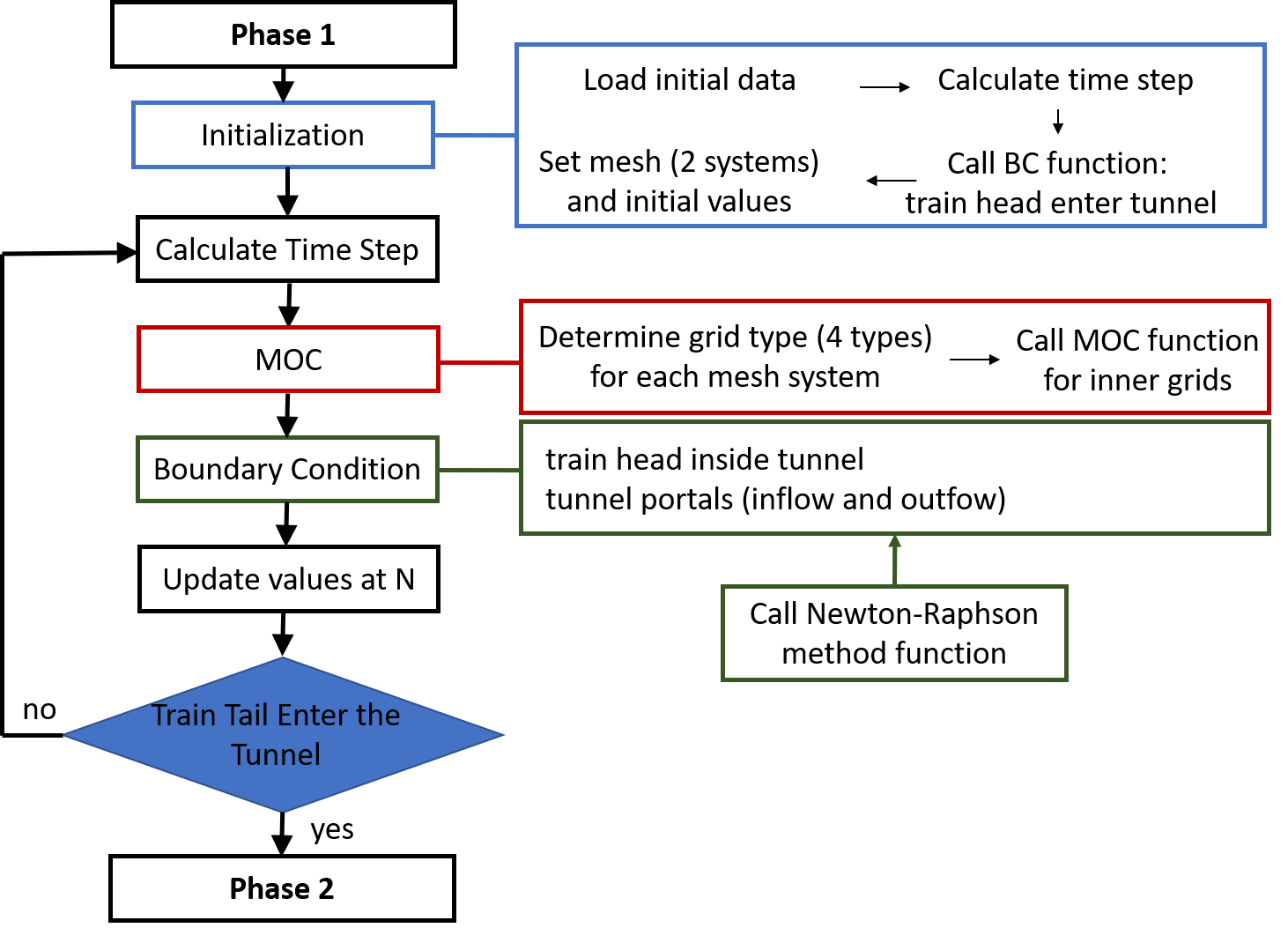}
    \caption{\textcolor{black}{Flow chart of Phase 1, illustrating the process from the entry of the train head into the tunnel to the entry of the train tail.}}
    \label{fig:Flow chart}
\end{figure}

The entire programme is constructed using three subroutines, depending on the train's relative position to the tunnel. The programme flow chart of phase 1 (from train head entering the tunnel to train tail entering the tunnel) is shown as Figure.~\ref{fig:Flow chart}. For each subroutine, the calculation process is similar. First, is to judge the type of the mesh and obtain the characteristic value at moment $Z$ on every grid point. Then calculate the time step $\Delta Z$, choosing the minimum $\Delta Z$ to keep it consistent among all the ducts. The time step is calculated based on Courant–Friedrichs–Lewy (CFL) criterion $\frac{\Delta Z}{\Delta X}\leq\frac{1}{A+|U|}$ (where $\Delta X$ refers to grid spacing, \textcolor{black}{$A$ and $U$ denote the non-dimensional speed of sound and gas velocity, respectively, both normalized by the reference speed of sound in air.}). Then by considering the direction of velocity, determine the position of interpolated points and further calculate the characteristic value of all the inner points at $Z+\Delta Z$. The characteristic value of boundary grids are calculated according to the corresponding boundary conditions. The equations for boundary conditions at the tunnel portals and the train's head and tail are provided in Liu et al. \cite{liu2023study}. Additionally, the boundary conditions at the discontinuities caused by the large gaps in the partially loaded freight train are derived in Section~\ref{chap:BC}. Then the pressure and velocity can be obtained based on the relationship between Riemann variables and thermodynamics. Finally, assign the value of  $Z+\Delta Z$ to  $Z$, and execute the loop until the phase end.

\subsubsection{Mesh}
\begin{figure}[htbp]
    \centering   
    \includegraphics[width=12cm]{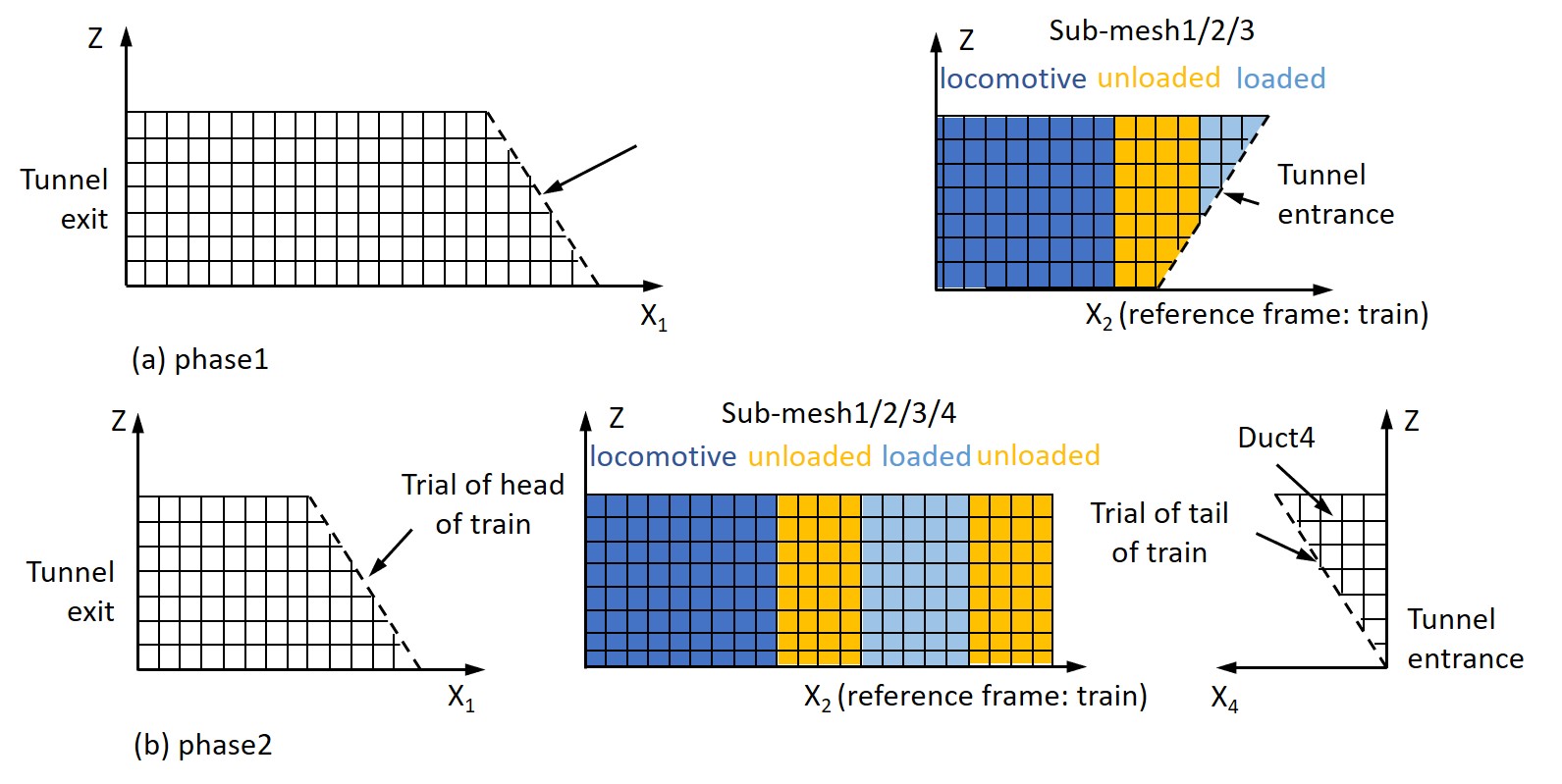}
    \caption{Modified mesh system for a partially loaded container freight train.}
    \label{fig:mesh_partially_loaded}
\end{figure}

 The modified mesh system of a locomotive with a partially loaded container freight train entering a tunnel is developed, as illustrated in Figure~\ref{fig:mesh_partially_loaded}, in which the grid in the annulus space ($X_2$) is based on the train as the frame of reference. The annulus space around the train is further divided into sub-meshes by the ends of the gap between containers. The flow in each sub-mesh system is solved using the MOC method based on the corresponding different physical parameters (length, perimeter, cross-sectional area, etc.), where the flow connecting different sub-meshes is solved by the boundary conditions illustrated in Section~\ref{sec:1D_BC}. Through this approach, it can be seen from Figure~\ref{fig:mesh_partially_loaded} that although there are discontinuities along the train body, it doesn't have to separate the mesh at the annulus space when the train enters the tunnel, which avoids the excessive extending and extracting the mesh, as would be the case if the mesh frame of reference was with respect to the static ground, rather than the train as is the case in this study. 

\subsubsection{Boundary conditions} \label{sec:1D_BC}
\begin{figure}[htbp]
    \centering   
    \includegraphics[width=16cm]{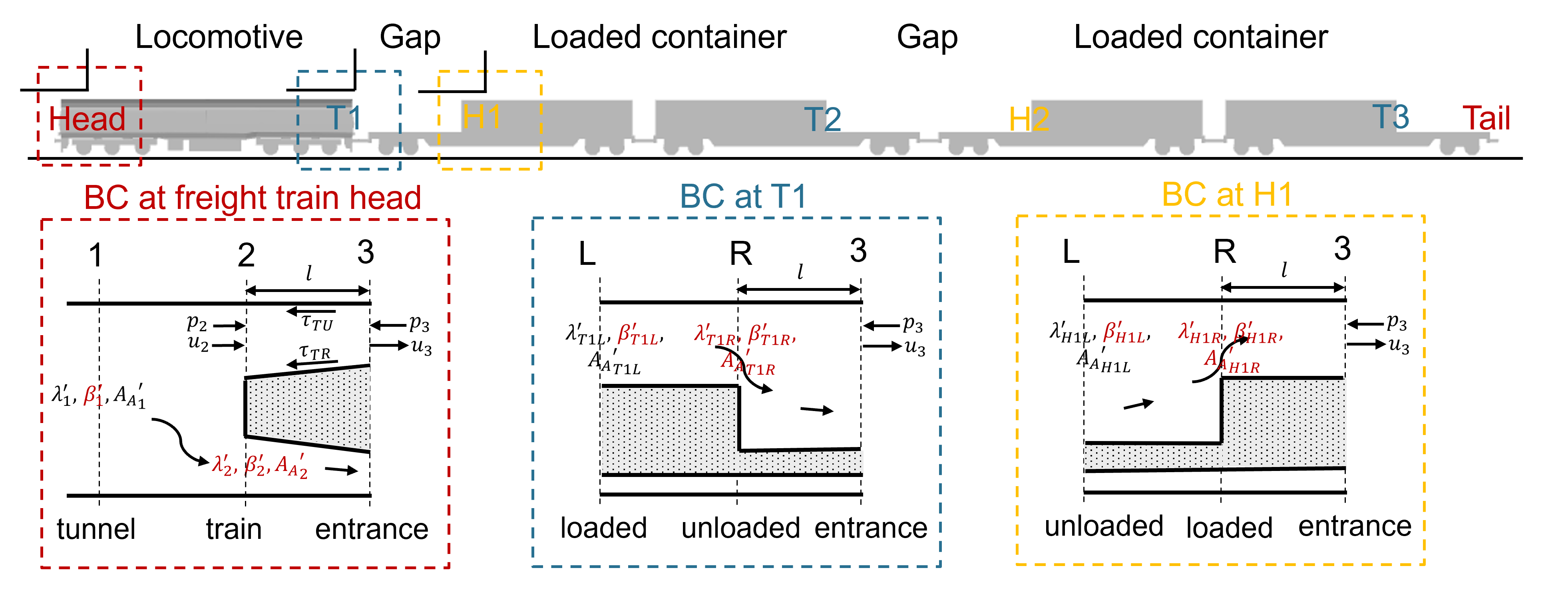}
    \caption{\textcolor{black}{New boundary condition at the ends of containers for an example partially loaded container freight train (BC: Boundary condition).}}
    \label{fig:BC_partially_loaded}
\end{figure}

The boundary conditions at the stationary mesh end, and the tunnel portals, are solved using the traditional method as presented by Woods et al. \cite{woods1981generalised}. Additionally, a new approach to setting up the boundary conditions at the moving mesh end, which specifically relates to the head and tail of the locomotive, considers the effective area change caused by the presence of separation bubbles. For the completeness of this paper, a summary of the boundary condition of solving the Riemann variables ($\lambda$,$\beta$,$A_A$) for the train head entering the tunnel is presented as Equations ~\ref{eq:head_equation1} to \ref{eq:head_equation4}, \textcolor{black}{and the corresponding schematic to understand these equations are explained in the sub-figure 'BC at freight train head' in Figure.~\ref{fig:BC_partially_loaded}. Subscripts 1 and 2 represent the Riemann variables at two adjacent boundary points—one inside the tunnel and one at the train head—at the current time step. Subscript 3 represents the point at tunnel portal.}
\begin{equation}
\left(\lambda_1^{\prime}+\beta_1^{\prime}\right)^2-\left(\lambda_2^{\prime}+\beta_2^{\prime}\right)^2+\frac{2}{\kappa-1}\left[\left(\lambda_1^{\prime}-\beta_1^{\prime}\right)^2-\left(\lambda_2^{\prime}-\beta_2^{\prime}\right)^2\right]=0
\label{eq:head_equation1}
\end{equation}
\begin{equation}
\left(\lambda_1^{\prime}-\beta_1^{\prime}\right)^2\left(\lambda_1^{\prime}+\beta_1^{\prime}\right)^{\frac{2}{\kappa-1}}\left(\frac{A_{A_2}}{A_{A_1}}\right)^{\frac{2 \kappa}{\kappa-1}}-\frac{E_2}{E_1}\left(\lambda_2^{\prime}-\beta_2^{\prime}\right)\left(\lambda_2^{\prime}+\beta_2^{\prime}\right)^{\frac{2}{\kappa-1}}=0
\label{eq:head_equation2}
\end{equation}
\begin{equation}
\frac{A_{A_2}}{A_{A_1}}=\left( 1 + \frac{\frac{2 \kappa}{(\kappa-1)^2}\left(\frac{\lambda_2^{\prime}-\beta_2^{\prime}}{\lambda_2^{\prime}+\beta_2^{\prime}}\right)^2 \zeta_N}{\left[1+\frac{2}{\kappa-1}\left(\frac{\lambda_2^{\prime}-\beta_2^{\prime}}{\lambda_2^{\prime}+\beta_2^{\prime}}\right)^2\right]^{\frac{\kappa}{\kappa-1}}} \right)^ {\left( \frac{\kappa - 1}{2 \kappa} \right)}
\label{eq:head_equation3}
\end{equation}
\begin{equation}
\begin{aligned}
\left(\frac{\lambda_2^{\prime}+\beta_2^{\prime}}{2 A_{A_2}}\right)^{\frac{2 \kappa}{\kappa-1}}= & \frac{1}{p_R}\left\{\frac{p_3}{\phi}+\rho_3 c_R^2(\phi-1)\left(\frac{\lambda_2^{\prime}-\beta_2^{\prime}}{\kappa-1}+\frac{V}{c_R}\right)^2\right. \\
& +\frac{\rho_3 l c_R^2}{2 E_2}\left[\frac{f_{\textcolor{black}{tu}} S_{\textcolor{black}{tu}}}{4}\left[(\phi+1)\left(\frac{\lambda_2^{\prime}-\beta_2^{\prime}}{\kappa-1}+\frac{V}{c_R}\right)\right]^2\right. \\
& \left.\left.+\frac{f_{\textcolor{black}{tr}} S_{\textcolor{black}{tr}}}{4}\left[(\phi+1)\left(\frac{\lambda_2^{\prime}-\beta_2^{\prime}}{\kappa-1}\right)\right]^2\right]\right\}
\end{aligned}
\label{eq:head_equation4}
\end{equation}
Here $E$ and $S$ represent the area and perimeter respectively.
Now, if we consider an intermodal container freight train that has a  partially loaded container configuration, there are potentially large gaps between these loaded containers which are likely to be regions that induce a pressure loss. Therefore, extending the idea presented for the locomotive boundary conditions, there is a need to consider both ends of these large gaps, which again can be considered as discontinuities in the 1D code. Therefore, new boundary conditions at the front and end of these gaps entering the tunnel are added to the code to describe these discontinuities, as illustrated in Figure~\ref{fig:BC_partially_loaded}. The air flow from the loaded/unloaded region to the unloaded/loaded region can be simplified as a duct flow from large/small cross-sectional area, to the small/large cross-sectional area. In summary, the flow characteristic at these positions can be considered as ducts with a sudden area change. This is a similar concept to that applied to simulate the flow at locomotive head and tail. The relationship of velocity, pressure and density between connected ducts and the deduction of continuity, momentum and energy equation for quasi-steady flow is similar as those illustrated in previous papers \cite{woods1981generalised}. Thus in this section, only the newly developed equations used to directly solve the unknown Riemann variables \textcolor{black}{when loaded and unloaded containers enter the tunnel} are presented,  as given in Equations~\ref{eq:partially_loaded_T1_equation1}-\ref{eq:partially_loaded_T1_equation3}. The unknown Riemann variables at these boundaries are marked red in Figure~\ref{fig:BC_partially_loaded} \textcolor{black}{(BC at T1 and BC at H1)}.
\begin{equation}
\left(\lambda_{T1L}^{\prime}+\beta_{T1L}^{\prime}\right)^2-\left(\lambda_{T1R}^{\prime}+\beta_{T1R}^{\prime}\right)^2+\frac{2}{\kappa-1}\left[\left(\lambda_{T1L}^{\prime}-\beta_{T1L}^{\prime}\right)^2-\left(\lambda_{T1R}^{\prime}-\beta_{T1R}^{\prime}\right)^2\right]=0
\label{eq:partially_loaded_T1_equation1}
\end{equation}
\begin{equation}
\left(\lambda_{T1L}^{\prime}-\beta_{T1L}^{\prime}\right)^2\left(\lambda_{T1L}^{\prime}+\beta_{T1L}^{\prime}\right)^{\frac{2}{\kappa-1}}\left(\frac{A_{A_{T1R}}}{A_{A_{T1L}}}\right)^{\frac{2 \kappa}{\kappa-1}}-\frac{E_{T1R}}{E_{T1L}}\left(\lambda_{T1R}^{\prime}-\beta_{T1R}^{\prime}\right)\left(\lambda_{T1R}^{\prime}+\beta_{T1R}^{\prime}\right)^{\frac{2}{\kappa-1}}=0
\label{eq:partially_loaded_T1_equation2}
\end{equation}
\begin{equation}
\frac{A_{A_{T1R}}}{A_{A_{T1L}}}=\left( 1 + \frac{\frac{2 \kappa}{(\kappa-1)^2}\left(\frac{\lambda_{T1R}^{\prime}-\beta_{T1R}^{\prime}}{\lambda_{T1R}^{\prime}+\beta_{T1R}^{\prime}}\right)^2 \zeta}{\left[1+\frac{2}{\kappa-1}\left(\frac{\lambda_{T1R}^{\prime}-\beta_{T1R}^{\prime}}{\lambda_{T1R}^{\prime}+\beta_{T1R}^{\prime}}\right)^2\right]^{\frac{\kappa}{\kappa-1}}} \right)^ {\left( \frac{\kappa - 1}{2 \kappa} \right)}
\label{eq:partially_loaded_T1_equation3}
\end{equation}
\textcolor{black}{Here $\lambda^{\prime}, \beta^{\prime}$, and $A_A$ are the Riemann variables relative to the train. Note that $E_{T1L}$ and $E_{T1R}$ are the areas of the left and right annular ducts after accounting for flow separation, which includes the boundary layer separation at the rear of the loaded container or locomotive, as well as the formation of a separation bubble near the leading edge of the container following the gap.
} $\zeta$ is the pressure loss coefficient at the discontinuities at the ends of the unloaded containers. The pressure loss at T1, $\zeta_{T1}$, is calculated based on the same method (Borda-Carnot relationship \citep{vardy1999estimation}) to calculate the pressure loss at train tail, using the effective cross-sectional areas of the loaded and unloaded containers,
\begin{equation}
\zeta_{T1}=(\frac{E_{con}-E_{gap}}{E_{tu}-E_{con}})^2
\end{equation}
in which $E_{con}$ is the cross-sectional area of the loaded container and \textcolor{black}{$E_{gap}$} is the effective cross-sectional area of the gap, \textcolor{black}{which is the section between two loaded containers}. Apart from the Equations ~\ref{eq:partially_loaded_T1_equation1}- \ref{eq:partially_loaded_T1_equation3}, an additional characteristic equation concerning $\beta_{T1L}$ is required to solve the four unknown variables. These equations are solved using Newton-Raphson method.

For the equations to solve the flow from the unloaded region to loaded region, the subscript T1 is substituted by H1, and the pressure loss at H1 is $\zeta_{H1}=0.03$. Furthermore, due to the sharp front edge of the container, a separation bubble forms when the air flows over from the gap in front. Thus apart from Equations~\ref{eq:partially_loaded_T1_equation1} - \ref{eq:partially_loaded_T1_equation3}, Equation~\ref{eq:head_equation5} is needed considering the area change caused by the separation bubble when each container enters the tunnel,  
\begin{equation}
\begin{aligned}
\left(\frac{\lambda_{T1R}^{\prime}+\beta_{T1R}^{\prime}}{2 A_{A_{T1R}}}\right)^{\frac{2 \kappa}{\kappa-1}}= & \frac{1}{p_R}\left\{\frac{p_3}{\phi}+\rho_3 c_R^2(\phi-1)\left(\frac{\lambda_2^{\prime}-\beta_2^{\prime}}{\kappa-1}+\frac{V}{c_R}\right)^2\right. \\
& +\frac{\rho_3 l c_R^2}{2 E_{T1R}}\left[\frac{f_{gap} S_{gap}}{4}\left[(\phi+1)\left(\frac{\lambda_{T1R}^{\prime}-\beta_{T1R}^{\prime}}{\kappa-1}+\frac{V}{c_R}\right)\right]^2\right. \\
& \left.\left.+\frac{f_{con} S_{con}}{4}\left[(\phi+1)\left(\frac{\lambda_{T1R}^{\prime}-\beta_{T1R}^{\prime}}{\kappa-1}\right)\right]^2\right]\right\}
\end{aligned}
\label{eq:head_equation5}
\end{equation}
Equation~\ref{eq:head_equation5} is derived from the quasi-steady flow in the short annular duct around the front of each container as it enters the tunnel. \textcolor{black}{$\phi$ is the annular area ratio between sections 'R' and '3', where the difference of the two areas arises due to the separation bubble formed near the front edge of the container. $V$ is the speed of the train. $l$ denotes the short distance by which the loaded container has entered the tunnel (as illustrated in Figure~\ref{fig:BC_partially_loaded}, boundary condition at H1). $E$ and $S$ represent the area and perimeter respectively, and $f$ is the frictional coefficient. }

\textcolor{black}{As can be seen from the equations in the modified 1D model used to calculate the unknown variables, certain input parameters describing the partially loaded freight train and the tunnel must be specified in advance, including atmospheric pressure and temperature; tunnel cross-sectional area, perimeter, and total length; tunnel surface friction parameter and portal pressure loss coefficient; train speed, cross-sectional area, perimeter, and length; train surface friction parameter and head/tail pressure loss coefficients; reference sound velocity, reference length, reference pressure, and grid size. It is important to note that the parameter values differ for the unloaded and loaded parts of the freight train. Furthermore, the size of the separation bubble changes with the size of the gap. Details of the relationship between these parameters which need to be determined and the influence of the gap size are discussed in the following sections.}

\subsection{3D simulations}
\textcolor{black}{This section presents the numerical setup for the 3D simulations conducted in this study. Two types of operating conditions are considered: one involving a train passing through the tunnel, and the other simulating a steady train inside the tunnel. The numerical model used to simulate the flow around the partially loaded freight train, along with the governing equations, is introduced in Section~\ref{chap:BC}. The corresponding boundary conditions for both operating conditions are described in Section~\ref{chap:mesh_3d}. Finally, the mesh resolution and numerical setup are validated against experimental data in Section~\ref{chap:model_validation}.}

\subsubsection{Numerical method} \label{chap:BC}
\textcolor{black}{Since the accurate prediction of pressure waves depends on capturing the separated vortices between containers, the wake vortex length, and the flow development in these regions, it is important to accurately resolve these features in the initial stage using 3D computational fluid dynamics techniques.} The study of these results will then enable key flow characteristics to be parametrised for the purpose of redeveloping the 1D numerical model.

Now when modelling the complex aerodynamic interaction of a train passing through a tunnel, it is known that the recirculation region features high negative wall shear stress along the train surface, which has the effect of decreasing the pressure rise caused by frictional effect. From previous studies, it is known that $k-\omega~sst$ URANS model tends to over predict the vortex reattachment length \cite{kalitzin2016improvements,liu2023study}, which leads to a substantial underestimation of the pressure rise caused by frictional effects during the train tunnel entry and an error on the flow development in the gap. Thus in order to accurately predict the pressure wave magnitude, all the 3D simulations carried out in this research are conducted using the LES numerical model, despite higher computational resources when compared to the URANS model.

A WALE (Wall modelled LES) model is adopted for the  subgrid-scale (SGS) model in this study. Previous research illustrates that the WALE model performs well in resolving complex, highly turbulent flows \citep{benchikh2019turbulent}, such as the flow created about the bluff freight train. The subgrid scale kinetic energy can be computed as follows, 
\begin{equation}
    \frac{\partial K_{t}}{\partial t} + \frac{\partial \Tilde{u}_{j} K_{t}}{\partial x_{i}} = \frac{\partial}{\partial x_{i}} \left(\frac{\mu_{t}}{\Bar{\rho} Pr_{t}} \frac{\partial K_{t}}{\partial x_{i}} \right) - \tau_{ij} \frac{\partial \Tilde{u}_{j}}{\partial x_{i}} - \epsilon_{t}
    \label{LES-WALE4}
\end{equation}
\textcolor{black}{where $K_t$ denotes the subgrid-scale turbulent kinetic energy, $Pr_t$ is the turbulent Prandtl number, and subscript $i$ refers to the spatial coordinate direction in index notation (i.e., $i = 1,2,3$ corresponds to $x$, $y$, and $z$).} $\epsilon_{t}$ and $\tau_{ij}$ represent the subgrid scale dissipation and stress tensor, and $\mu_{t}$ is the eddy-viscosity that is calculated based on WALE model,
\begin{equation}
    \mu_{t} = \rho L_{s}^{2} \frac{(S_{ij}^{d}S_{ij}^{d})^{3/2}}{(\Bar{S}_{ij}\Bar{S}_{ij})^{5/2} + (S_{ij}^{d}S_{ij}^{d})^{5/4}} 
    \label{LES-WALE5}
\end{equation}

\textcolor{black}{where $L_s$ is the local filter width, $S_{ij}$ denotes the filtered strain-rate tensor. A second-order implicit scheme is adopted for time discretisation, coupled with the PISO algorithm to handle pressure–velocity coupling. The time step is set to $4 \times 10^{-5}$~s, which satisfies the CFL condition ($CFL \leq 1$) based on the train speed and the minimum grid size (\SI{0.0015}{m}).} Furthermore, given that this study focuses on different container loading configurations for wagons, rather than just the freight locomotive, a supplementary simulation was conducted. This simulation was compared with experimental results to ensure that the mesh resolution in the gaps and around the containers is sufficiently fine to accurately simulate the flow field, as described in Section \ref{chap:model_validation}.

\subsubsection{Boundary Conditions and Mesh} \label{chap:mesh_3d}

\begin{figure}[H]
    \centering   
    \begin{subfigure}[t]{0.52\textwidth}
    \centering
   \includegraphics[width=\textwidth]{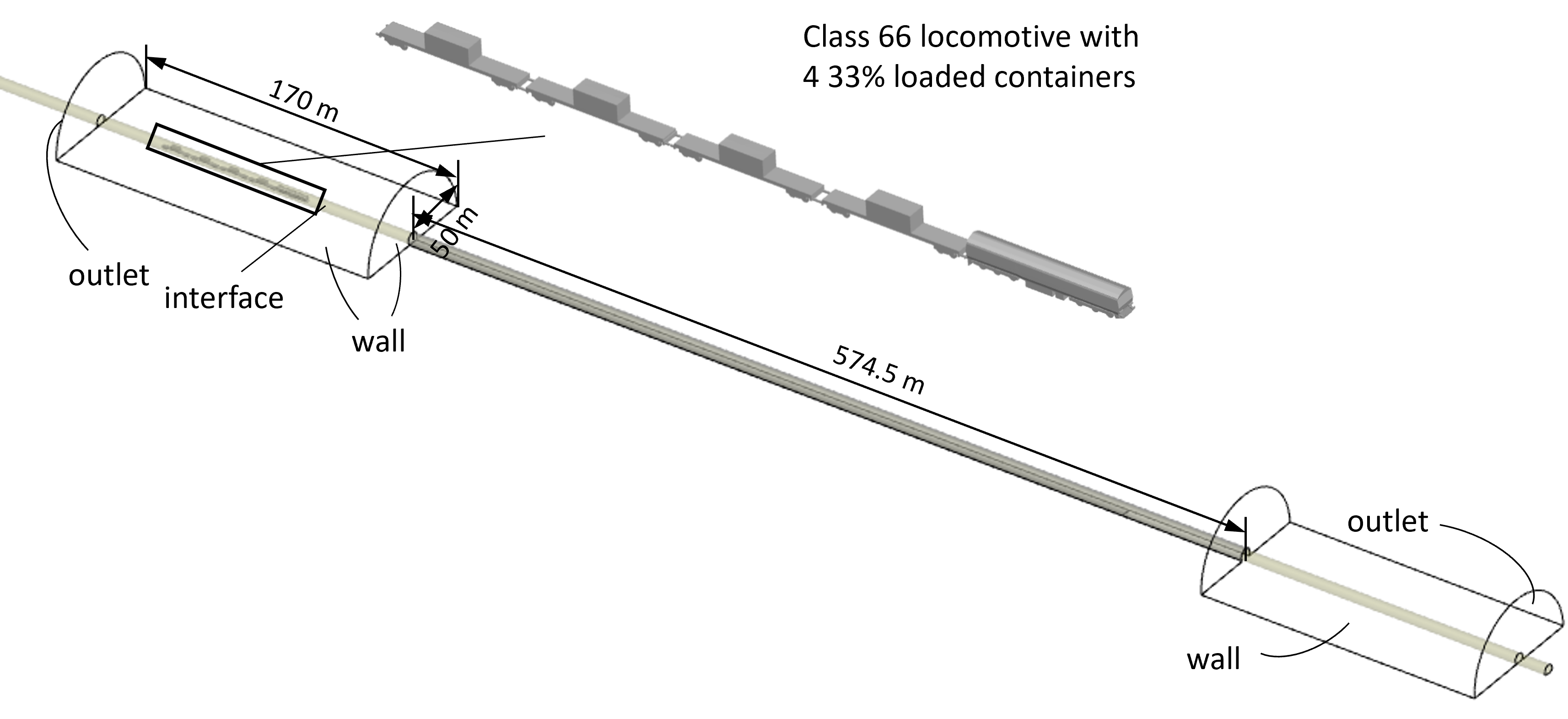}
   \caption*{\text{\small (a)}}
    \end{subfigure}
    \begin{subfigure}[t]{0.46\textwidth}
    \centering
   \includegraphics[width=\textwidth]{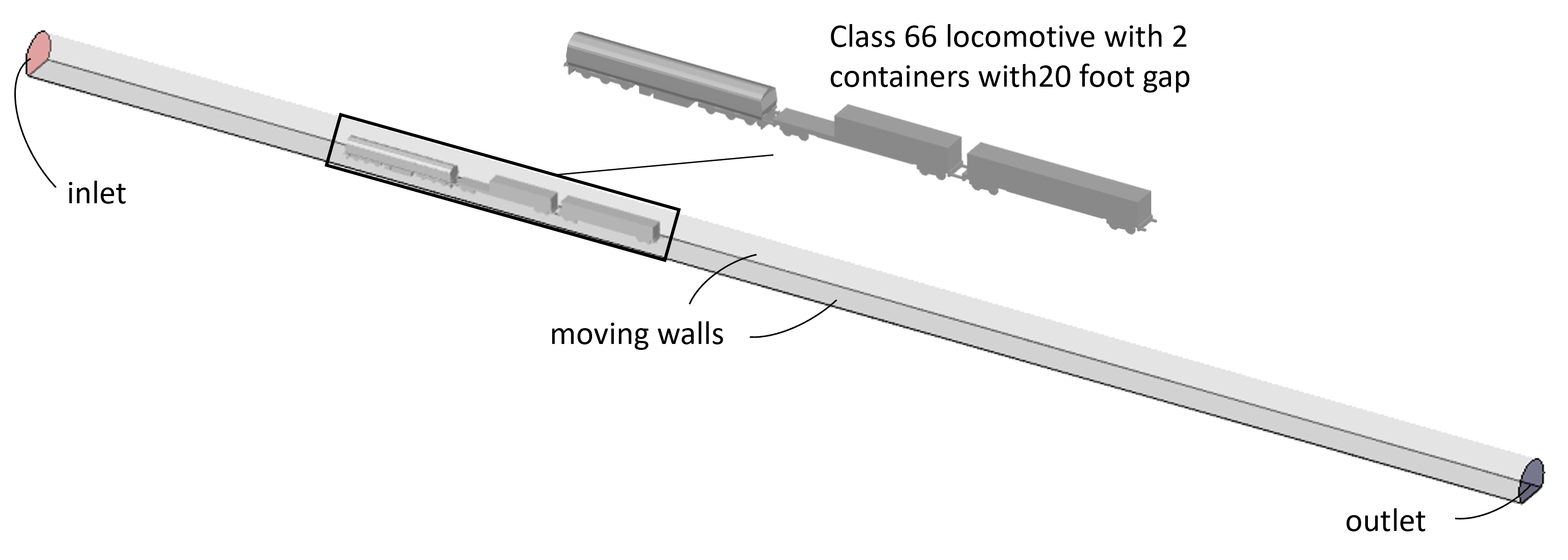}
   \caption*{\text{\small (b)}}
    \end{subfigure}
    \caption{Computational domain of 33\% loaded freight train passing through a tunnel and boundary conditions (dimensions in the figure are given as full-scale).}
    \label{fig:computational_domain1}
\end{figure}

This subsection presents the computational setup used in the 3D LES simulations. Two types of operating conditions are considered: one involving a train moving through the tunnel and the other simulating a steady train inside the tunnel. \textcolor{black}{For the transient case, a sliding mesh approach is employed to capture the relative motion between the train and tunnel accurately. Specifically, the freight train is embedded within a moving mesh region, which slides through a stationary tunnel mesh, as illustrated in Figure~\ref{fig:computational_domain1}(a).} In the parameterisation study of different loading conditions, the train is modelled as stationary inside the tunnel, while the tunnel walls and ground are simulated as moving walls to represent the relative motion between the train and tunnel as shown in Figure~\ref{fig:computational_domain1}(b).

% To accurately capture the flow separations around the containers downstream, the whole flow field is manually divided by structured mesh to control and improve the quality. The mesh blocking strategy and mesh resolution along the train body, including the containers and inter-container gaps is given in Figure~\ref{fig:mesh_blocking}. Specifically, 17 prism layers are deployed with $y^+ \approx 8$ near the walls to resolve the viscous sublayer, and mesh sizes in the streamwise and spanwise directions are based on best practices for LES of bluff-body flows \citep{osth2014study,krajnovic2005flowb}.

\begin{figure}[H]
    \centering   
    \begin{subfigure}[t]{0.75\textwidth}
    \centering
   \includegraphics[width=\textwidth]{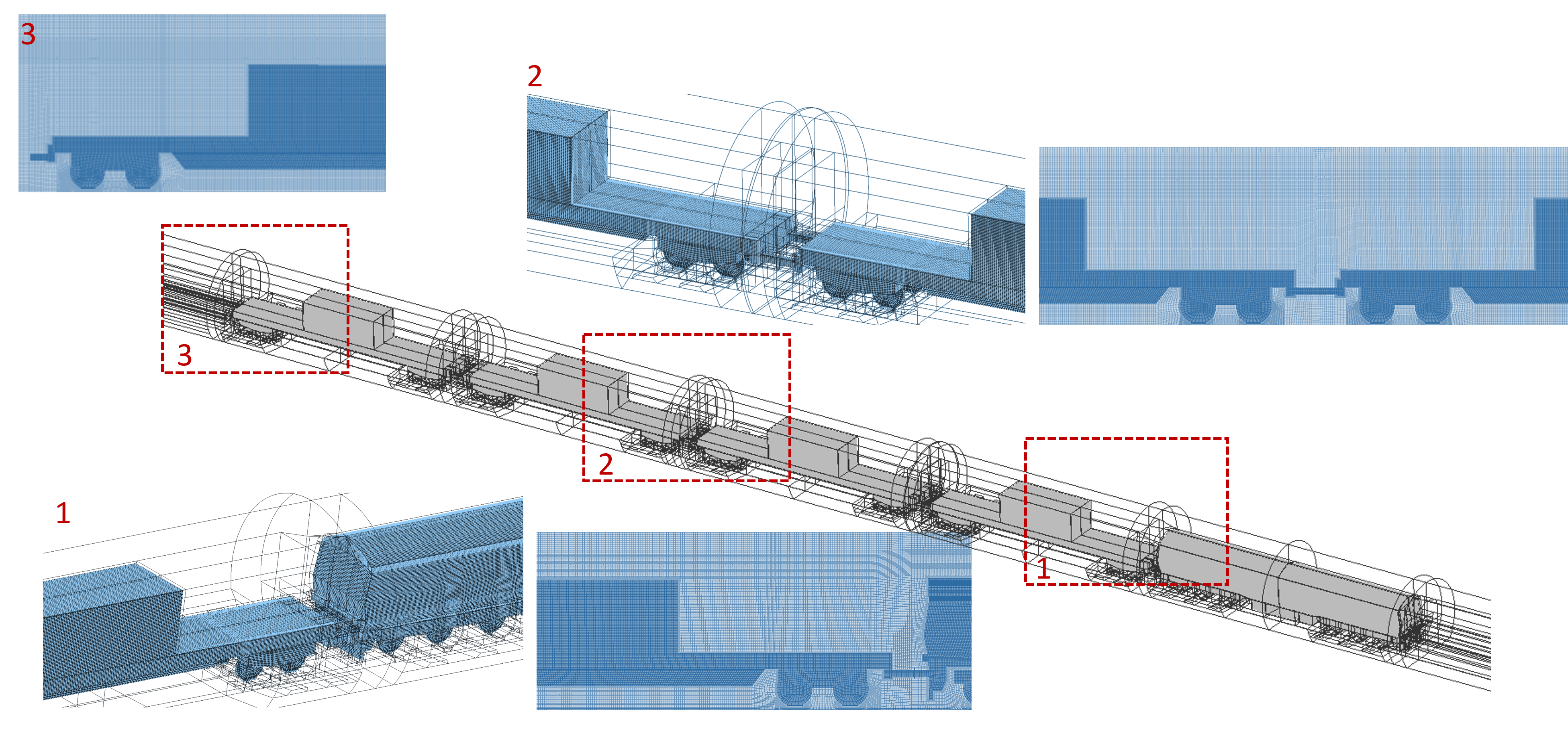}
    \end{subfigure}
    \caption{Blocking structure and mesh detail around the 33 $\%$  partially loaded freight train.}
    \label{fig:mesh_blocking}
\end{figure}

For the case of a freight train with a Class 66 locomotive hauling four wagons loaded with a 20-foot container in the centre of each, providing a loading efficiency of 33$\%$, the blocking structure and the mesh around the train are shown in Figure~\ref{fig:mesh_blocking}. Since the prediction of the boundary layer separation is highly sensitive to mesh quality and the chosen numerical model, the entire computational domain has been divided using a structured mesh to enhance overall mesh quality \citep{krajnovic2002large,iliadis2020numerical}. The domain is meticulously divided into hundreds of hexahedral block regions. Unlike automatically generated meshes controlled by predefined parameters, a manually crafted structured mesh allows for precise control over each element's quality. This approach minimises the occurrence of poor-quality meshes, particularly at sharp edges or areas with significant curvature changes. For the mesh shown in Figure~\ref{fig:mesh_blocking}, more than 99.5$\%$ of the mesh exhibits a skewness quality lower than 0.54. As a result, the number of cells for the reference case at the moving and the steady region are 42.2 million and 3.7 million respectively.

To accurately capture the flow separations around the containers downstream, it is crucial to correctly simulate the flow development along the train body.
% Therefore, the number and thickness of the prism layers and the size of the mesh along the train body, including the containers and the gaps, are consistent with the mesh settings detailed for the locomotive in \citep{liu2023study}. 
In summary, the train surface mesh is designed with $y^{+} \approx 8$ and 17 prism layers to simulate the velocity distribution in the viscous flow region effectively \textcolor{red}{\citep{hesse2021simulation}}. Mesh sizes in the spanwise and streamwise directions are adjusted based on prior research \citep{osth2014study,krajnovic2005flowb}. 

\subsubsection{Validation} \label{chap:model_validation}
\begin{figure}[H]
    \centering   
    \begin{subfigure}[t]{0.7\textwidth}
    \centering 
   \includegraphics[width=\textwidth]{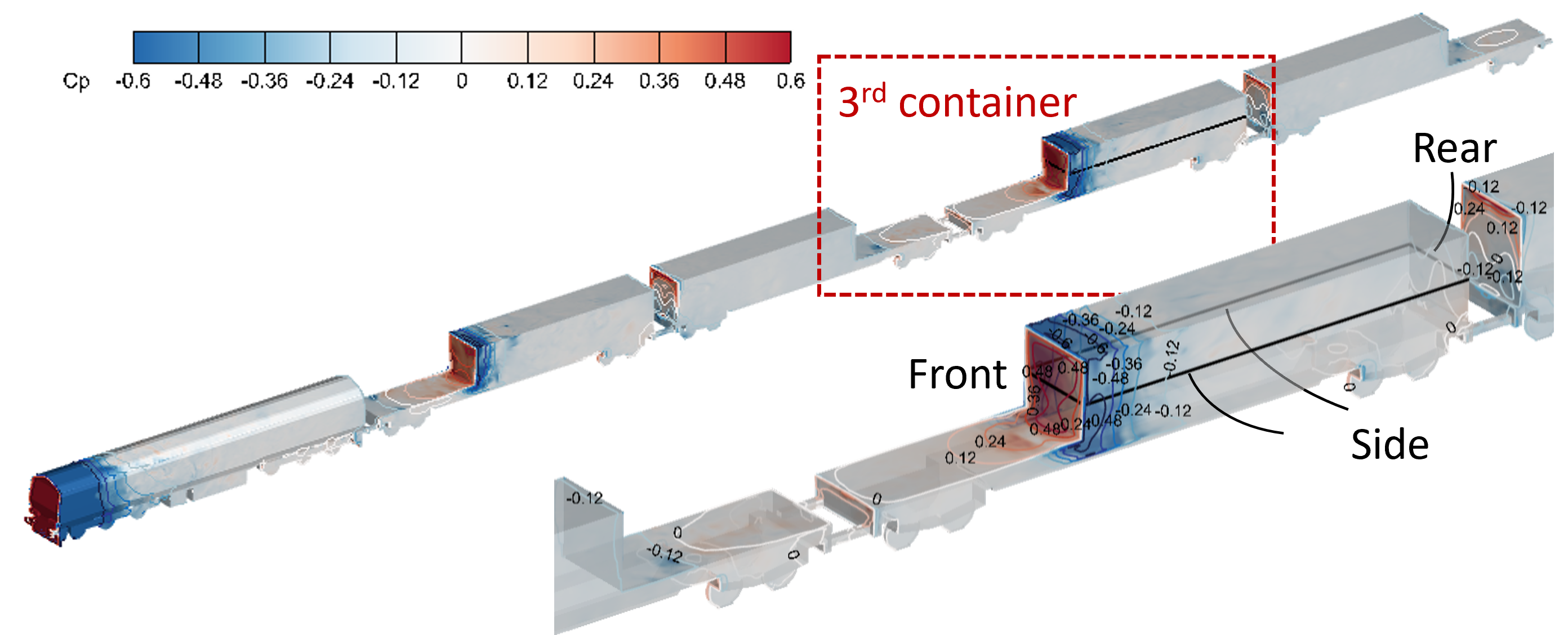}
    \caption*{\text{\small (a)}}
    \end{subfigure}
        \centering   
    \begin{subfigure}[t]{0.6\textwidth}
    \centering
   \includegraphics[width=\textwidth]{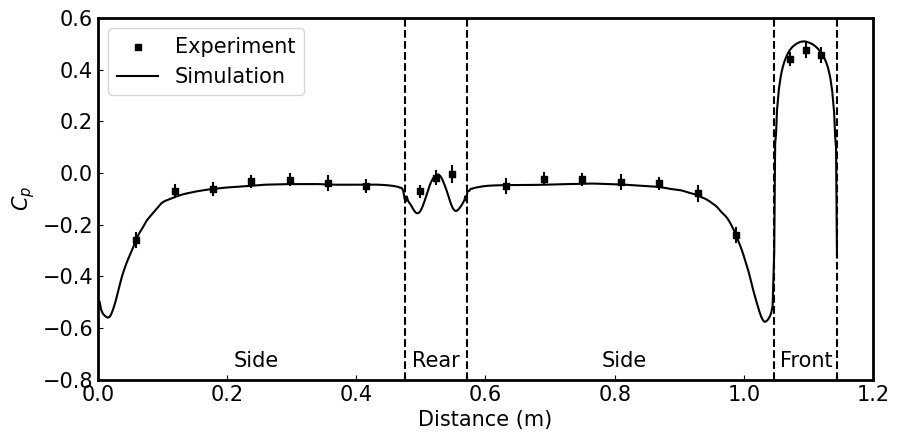}
    \caption*{\text{\small (b)}}
    \end{subfigure}
    \caption{Pressure distribution of the Class 66 hauled container freight train with four wagons loaded with containers in a 66 $\%$ loading efficiency, compared with experimental results \cite{soper2016aerodynamics} (a) on train surface (b) at mid-height of the third container.}
    \label{fig:pressure_validation}
\end{figure}

The ability of this mesh size to correctly predict the separation region around the freight locomotive has been validated and a mesh sensitivity test was conducted in the previous study by Liu et al \citep{liu2023study}. To further ensure that the mesh is sufficiently fine to resolve the flow field along the train body, a supplementary simulation was conducted using a Class 66 locomotive with four wagons loaded with containers in a 66$\%$ loading efficiency, as shown in Figure~\ref{fig:pressure_validation}. The pressure coefficient ($C_p = \frac{P-P_0}{0.5\rho v^2}$) at the mid-height of the container located on the third wagon behind the locomotive was compared with experimental results \cite{soper2016aerodynamics}. The geometry used for the validation case is shown in Figure~\ref{fig:pressure_validation}(a), with the train surface coloured by instantaneous pressure coefficient $C_p$ and lined by the mean pressure coefficient $\bar{C_p}$. 

Due to the gap before the third wagon, a relatively large positive pressure zone forms at the front face of the third container, with a maximum magnitude of approximately $\Bar{C_p} = 0.5$. A significant negative pressure area forms at the front of the side surface due to flow separation at the leading edge of the container. By quantitatively comparing the simulation results of $\Bar{C_p}$ at the mid-height of the container as shown in Figure~\ref{fig:pressure_validation}(b), it is evident that both the magnitude and position of the pressure zones on each surface of the third container fit well with the experimental results, with an error below 6.4$\%$. This indicates that the mesh is sufficiently fine to appropriately resolve the regions of separated flow. Moreover, it confirms that the upstream flow field, particularly relating to the gap between the second and third container, as well as the boundary layer separation at the rear of the containers located in the upstream flow field, has been accurately simulated.

\subsection{Numerical cases}
 Table~\ref{tab:Summary_cases_loadingconfiguration} has been constructed as a point of reference for the numerical simulations presented in the results and discussion sections.
 Case 1 is the reference case, by means to investigate the instantaneous flow pattern around the freight train of intermodal wagons with a series of containers in a  partially loaded formation. The flow characteristics are used for constructing the numerical model in the 1D programme. The 3D simulation results are then used to validate the feasibility of the modified 1D code by comparing the results of pressure history curve obtained from Case 2. Subsequently, Cases 3 to 5 focus on a parameterisation study of container loading configurations. This involves analysing how the unloaded gap impacts the effective cross-sectional area within the gap and the size of the separation bubble formed at the leading edge of the trailing container. To further validate the proposed model and the parameterisation equations, the insights gained from this parameterisation study are then incorporated into the 1D programme (Case 7) to simulate the pressure wave generated when a freight train enters a tunnel, where the first carriage is fully empty and the second carriage is 33 $\%$ loaded. A corresponding LES simulation (Case 6) is also carried out to compare with the results from the 1D code and validate the applicability of the proposed model. 

\begin{table}[H]
  \footnotesize
  \centering
  \begin{tabular}{|c|c|c|c|c|c|}
    \hline       
    \textcolor{black}{\textbf{Case}} & \textcolor{black}{\textbf{Train geometry}} & \textcolor{black}{\textbf{Tunnel}} & \textcolor{black}{\textbf{Operating condition}} & \textcolor{black}{\textbf{Model}} & \textcolor{black}{\textbf{Purpose}} \\
    \hline
    \textcolor{black}{1} & \textcolor{black}{4 wagons, 33\% loaded} & \textcolor{black}{45 m$^2$} & \textcolor{black}{Tunnel entry} & \textcolor{black}{1D model} & \textcolor{black}{Reference case} \\
    \textcolor{black}{2} & \textcolor{black}{4 wagons, 33\% loaded} & \textcolor{black}{45 m$^2$} & \textcolor{black}{Tunnel entry} & \textcolor{black}{LES} & \textcolor{black}{1D model validation} \\
    \textcolor{black}{3} & \textcolor{black}{2 containers, 20-foot gap} & \textcolor{black}{45 m$^2$} & \textcolor{black}{Steady inside tunnel} & \textcolor{black}{LES} & \textcolor{black}{Parameterisation} \\
    \textcolor{black}{4} & \textcolor{black}{2 containers, 40-foot gap} & \textcolor{black}{45 m$^2$} & \textcolor{black}{Steady inside tunnel} & \textcolor{black}{LES} & \textcolor{black}{Parameterisation} \\
    \textcolor{black}{5} & \textcolor{black}{2 containers, 60-foot gap} & \textcolor{black}{45 m$^2$} & \textcolor{black}{Steady inside tunnel} & \textcolor{black}{LES} & \textcolor{black}{Parameterisation} \\
    \textcolor{black}{6} & \textcolor{black}{1st wagon empty, 2nd 33\% loaded} & \textcolor{black}{70 m$^2$} & \textcolor{black}{Tunnel entry} & \textcolor{black}{LES} & \textcolor{black}{Validation} \\
    \textcolor{black}{7} & \textcolor{black}{1st wagon empty, 2nd 33\% loaded} & \textcolor{black}{70 m$^2$} & \textcolor{black}{Tunnel entry} & \textcolor{black}{1D model} & \textcolor{black}{Validation} \\
    \hline
  \end{tabular}
  \caption{\textcolor{black}{Summary of numerical cases used for the numerical model construction and validation.}}
  \label{tab:Summary_cases_loadingconfiguration}
\end{table}

\section{Numerical Results}
\label{chap:Numerical_Results}
\subsection{Reference case} \label{chap:reference_case}
\subsubsection{Pressure history curve}
\begin{figure}[htbp]
    \centering
    \includegraphics[width=10cm]{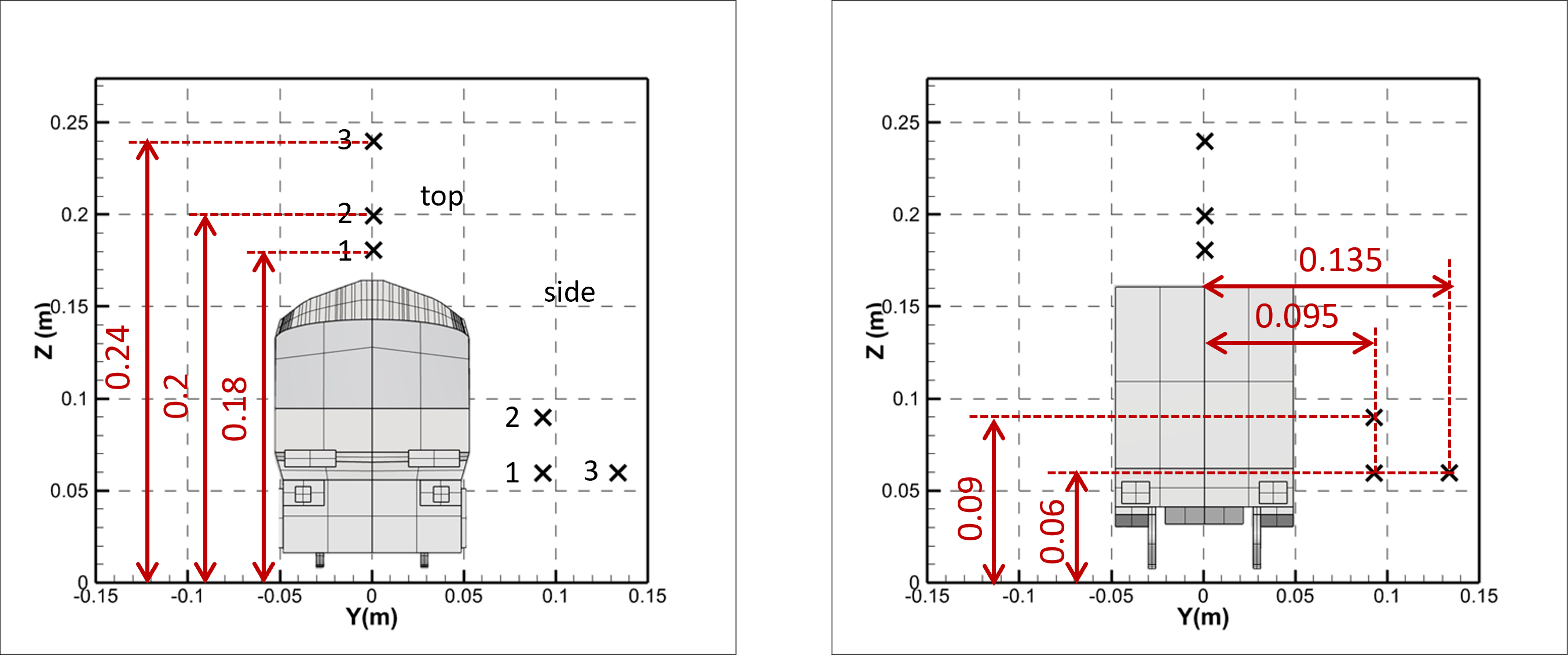}
    \caption{\textcolor{black}{Position of measuring lines relative to the locomotive and container (Dimensions are given in 1:25 reduced scale).}}
    \label{fig:line_positions}
\end{figure}

\begin{figure}[htbp]
    \centering
    \includegraphics[width=10cm]{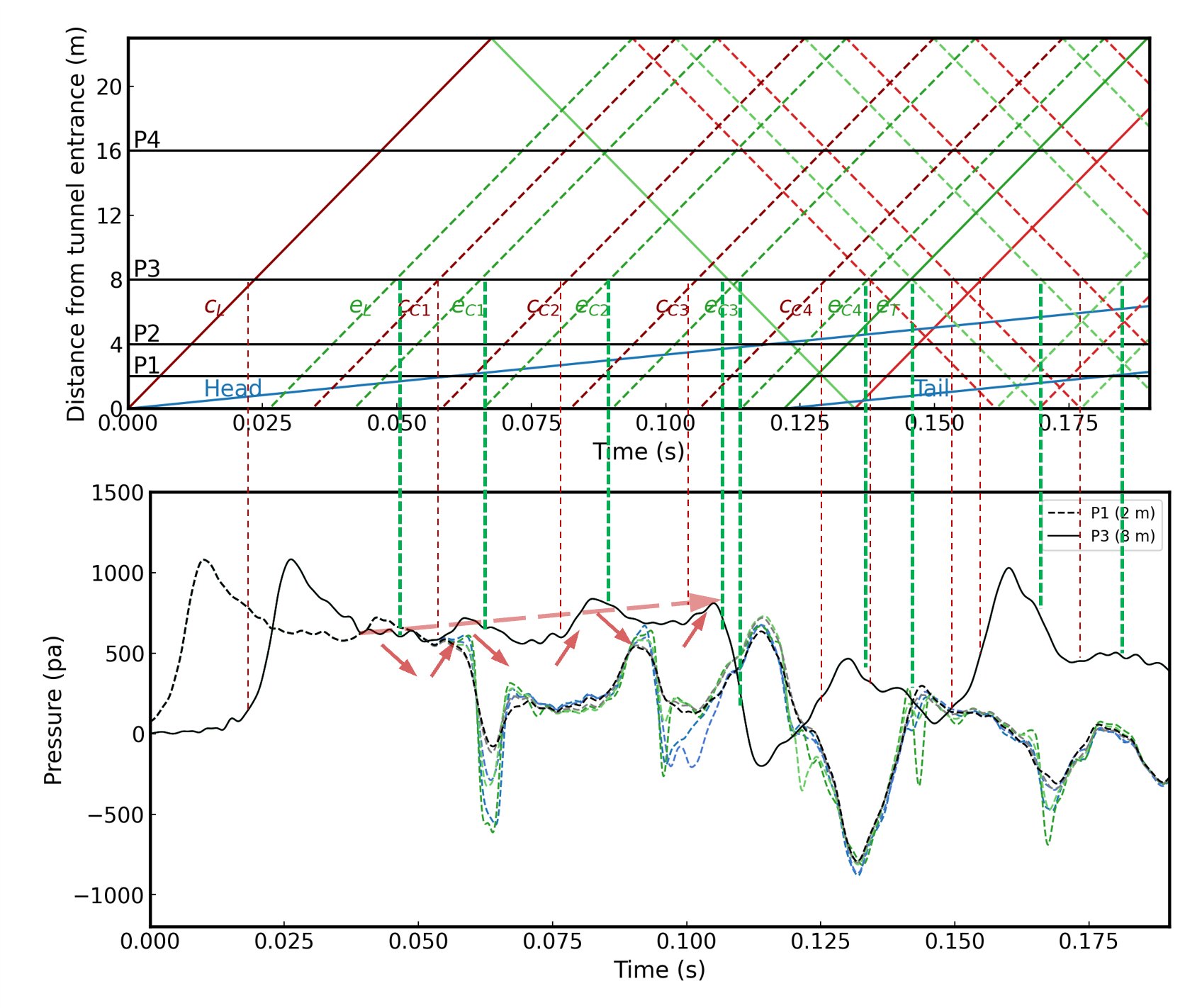}
    \caption{\textcolor{black}{Pressure history curve generated by a locomotive with four $33 \%$ partially loaded freight train at 2 m (solid lines) and 8 m (dashed lines) inside tunnel, each position have 6 measuring points ($c_{L}$: the first compression wave induced by the train head entering the tunnel;$e_L$: Expansion wave generated by the rear of the locomotive entering the tunnel; $c_{C1} \sim c_{C4}$: Compression wave generated by the front of the containers entering the tunnel; $e_{C1} \sim e_{C4}$: Expansion wave generated by the rear of the containers entering the tunnel; $e_T$: Expansion wave generated by the tail of the train entering the tunnel).}}
    \label{fig:diagram_pressurewaves}
\end{figure}

 In this section, a locomotive hauling four  $33 \%$ loaded intermodal freight wagons passing through a tunnel is simulated using the LES method and studied as the reference case. The pressure time-history curves are measured at locations P1 and P3 (2 m and 8 m respectively) inside the tunnel, and are shown in Figure~\ref{fig:diagram_pressurewaves}. At each location, six points are measured and with positions relative to train body illustrated in Figure.~\ref{fig:line_positions}. In order to clearly visualise the influence of pressure waves and the effect of the freight train body on the pattern of pressure history curves, Figure~\ref{fig:diagram_pressurewaves} is presented, showing the generation and propagation of the pressure waves and the track of train passing. Red lines represent compression waves, and the green lines represent expansion waves. Additionally, the dashed lines represent the sub-pressure waves generated by the containers on the intermodal flatbed wagons entering the tunnel. To explain the phenomenon, the sub-compression waves and expansion waves generated by the four partially loaded containers are labelled as $c_{C1}$ through $c_{C4}$ and $e_{C1}$ through $e_{C4}$, respectively. The trace lines of the head and tail of the whole freight train are marked as blue lines in the upper plot in Figure~\ref{fig:diagram_pressurewaves}. Additionally, the four measurement locations inside the tunnel (P1 - P4) at 2 m, 4 m, 8 m and 16 m \textcolor{black}{(corresponding to 50, 100, 200, and 400 m, respectively when converted to full-scale)} from the tunnel entrance are marked as black lines in Figure~\ref{fig:diagram_pressurewaves}. The interaction between these black lines and the pressure wave lines, as well as the trace lines of head and tail contributes to the pressure change on the pressure history curves at P1 and P3. \textcolor{black}{It can be observed that at the 8 m position (P3: solid lines), where the train has not yet reached the section, the pressure history curves within the tunnel completely overlap, indicating that the pressure changes induced by the incident waves are uniform across the cross section. In contrast, at the 2 m position (P1: dashed lines), the pressure varies among different measuring points as the train body passes by, revealing significant three-dimensional effects—particularly near the front and rear of the locomotive and each loaded container, as discussed in the previous section. This also highlights a limitation of current 1D models, which are unable to capture spatial variations in pressure along individual tunnel cross sections. Such discrepancies are more pronounced in partially loaded freight trains and warrant further investigation in future studies.}

\textcolor{black}{Moreover, the simulation results reveal that these pressure waves collectively contribute to an overall step-like pressure rise, as indicated by the red arrows in the figure. Specifically, the compression waves generated by the entry of the locomotive and the containers following the gap lead to successive increases in pressure, while the expansion waves generated by the tail of the locomotive—just before the gap—cause corresponding pressure decreases.} The initial pressure rise caused by $c_{C1}$ is $\sim$ 119 Pa, which is smaller than the subsequent pressure rises caused by $c_{C2}$ to $c_{C4}$ ($\sim$ 254 Pa). Similarly, the first pressure drop, caused by the end of the locomotive entering the tunnel ($e_{L}$) is $\sim$ 43 Pa, which is smaller than the subsequent pressure drops ($\sim$ 77 Pa). This occurs because the initial gap between the locomotive and the first container is not long enough for the flow to fully develop, leading to air becoming trapped in the gap and moving along with the train. This phenomenon increases the effective blockage area at the gap between containers, thereby decreasing the magnitude of the expansion waves as the tail of each container enters the tunnel. Furthermore, as more air becomes trapped in the gap, the separation bubble formed at the trailing edge of the container is smaller, reducing the strength of the compression wave formed when the trailing container enters the tunnel. Changes in the flow field around containers as the partially loaded freight train enters the tunnel are visualised and analysed in Section~\ref{chap:flow field}.


\subsubsection{Flow field} \label{chap:flow field}
\begin{figure}[H]
    \centering   
    \begin{subfigure}[t]{0.4\textwidth}
    \centering
    \includegraphics[width=\textwidth]{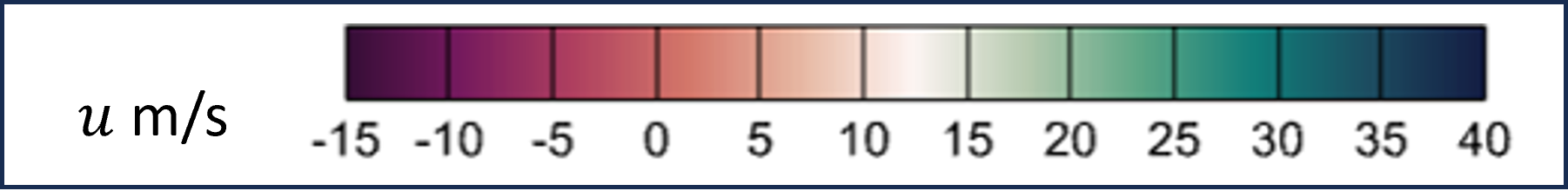}
    \end{subfigure}
    \centering   
    \begin{subfigure}[t]{0.45\textwidth}
    \centering
    \includegraphics[width=\textwidth]{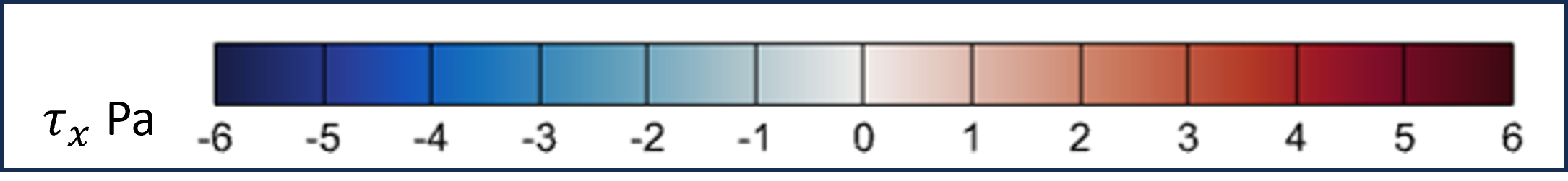}
    \end{subfigure}
    \centering   
    \begin{subfigure}[t]{0.9\textwidth}
    \centering
    \includegraphics[width=\textwidth]{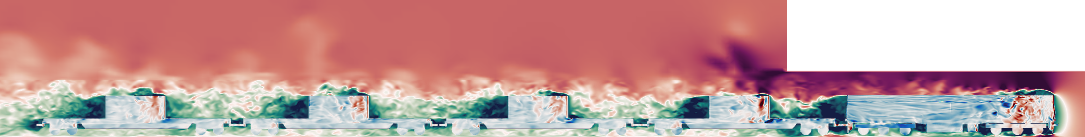}
    \caption*{\text{\small (a)}}
    \end{subfigure}
    \centering   
    \begin{subfigure}[t]{0.9\textwidth}
    \centering
    \includegraphics[width=\textwidth]{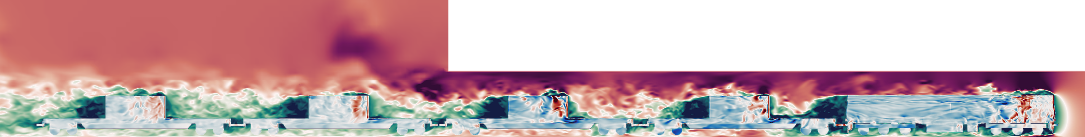}
    \caption*{\text{\small (b)}}
    \end{subfigure}
     \centering   
    \begin{subfigure}[t]{0.9\textwidth}
    \centering
    \includegraphics[width=\textwidth]{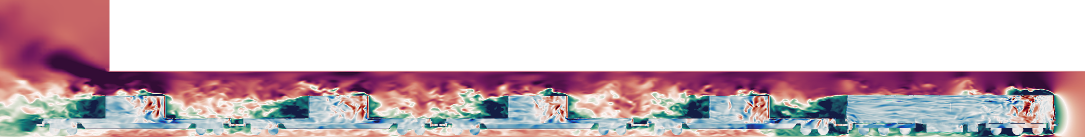}
    \caption*{\text{\small (c)}}
    \end{subfigure}
    \caption{Velocity contour at plane Y = 0 and distribution of surface wall shear stress at (a) t = 0.03 s (b) t = 0.07 s (c) 0.11 s.}
    \label{fig:flow_field}
\end{figure}

The flow characteristics around a partially loaded freight train influence the formation of pressure waves and are crucial for determining the parameters used in the proposed 1D model. To be more specific, when the main part of the train body enters the tunnel, the amplitude of the initial pressure wave rises, due to the frictional effect \cite{en140672021railway}, denoted as $\Delta p_{fr}$. In the case of passenger trains or fully laden freight trains, the train body maintains a streamlined form, facilitating the steady development of the boundary layer along its surface \citep{baker2001slipstream}. Consequently, the wall shear stress within this boundary layer is uniformly oriented, actively contributing to the pressure amplification caused by the pressure wave. However, for partially loaded container freight trains, a distinctive scenario unfolds due to the boundary layer separation after each container and the consequent formation of a recirculation zone at the forefront of each blunt container. To provide a comprehensive visualisation and analytical insight into the flow field, the instantaneous wall shear stress on the train body and the velocity contour during the process of a partially loaded freight train entering a tunnel are depicted in Figure~\ref{fig:flow_field}, from a side-view perspective.

The positive wall shear stress, denoted by the red colour, prominently characterises regions of recirculation or separation. The positive direction of the x-axis corresponds to the direction of the train's motion. As anticipated, the sections preceding each container and the inter-container gap exhibit this red hue, symbolising areas of notable recirculation and separation. The reversal in the direction of the wall shear stress within these separation regions affects the frictional dynamics and thereby decreases the pressure change $\Delta p_{fr}$. Upon comparative analysis of Figure \ref{fig:flow_field} (a) and (c), it is evident that the blue and red regions on the train body appear darker when inside the tunnel, implying a higher magnitude of wall shear stress in comparison to when outside the tunnel. This disparity can be attributed to the increased freestream velocity inside the confined tunnel space, resulting in an elevated friction velocity $u_\tau$. Moreover, the extensive coverage of red across the train body visually underscores that, in comparison to passenger trains, the impact of the frictional effect on the pressure change, $\Delta p_{fr}$, is relatively diminished for partially loaded freight trains. This explains the lower frictional coefficient adopted in the 1D code for partially loaded freight train. 

The side view of the instantaneous velocity contour at plane Y = 0 during the freight train entry into the tunnel is also depicted in Figure~\ref{fig:flow_field}. The negative $u$ values shown in the figure indicate flow directed from the train head toward the tail. Therefore, the green-shaded areas indicate recirculation zones. Consistent with the phenomenon observed from the distribution of wall shear stress, these zones are situated in the wake region of the leading container, at the front of the trailing container and around the former part of the containers. The separation bubbles formed at the top of the containers are notably smaller than those generated around the locomotive. This disparity arises because the trailing containers reside in the near or far wake region of the leading container, where the incoming air velocity is lower than the operating velocity of the train. Furthermore, the approaching air velocity in front of the trailing container increases with a rise in the gap distance. This phenomenon is similar to that observed in a platoon of vehicles \citep{vegendla2015investigation}. The dynamics of airflow within the gap significantly impact the effective blockage area of the unloaded section of the train and play a crucial role in simulating the 'step-like' pattern of the initial pressure wave, as shown in Figure~\ref{fig:diagram_pressurewaves}. However, due to the transient and turbulent nature of the flow calculated from the LES method, accurately determining the shape of the recirculation region is not possible. In Section~\ref{chap: parameterisationStudy_loadingconfiguration}, a quantitative analysis of this area will be carried out by analysing the mean flow field of a partially loaded train inside tunnel (train is stationary with velocity inlet at the tunnel entrance) using the LES method.

\begin{figure}[htbp]
    \centering   
    \begin{subfigure}[t]{0.3\textwidth}
    \centering
    \includegraphics[width=\textwidth]{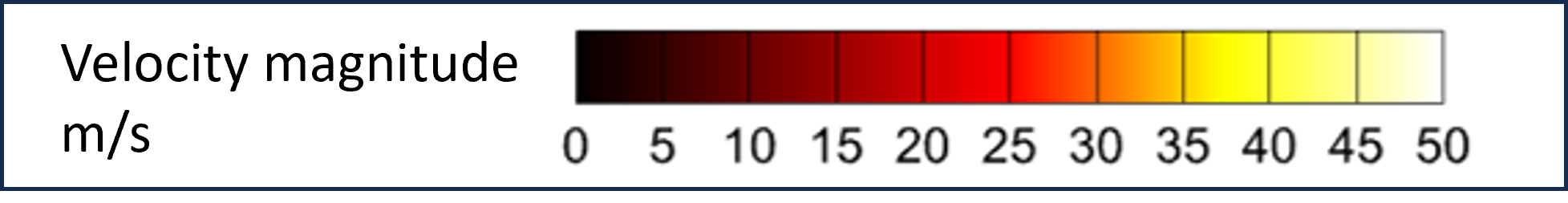}
    \end{subfigure}
    \centering   
    \begin{subfigure}[t]{0.3\textwidth}
    \centering
    \includegraphics[width=\textwidth]{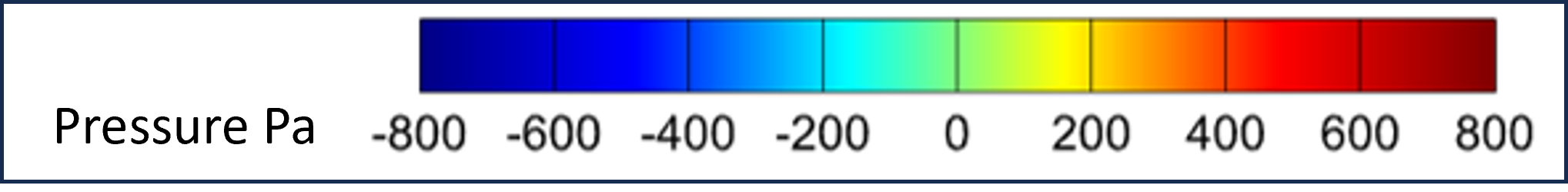}
    \end{subfigure}
    \centering   
    \begin{subfigure}[t]{0.9\textwidth}
    \centering
    \includegraphics[width=\textwidth]{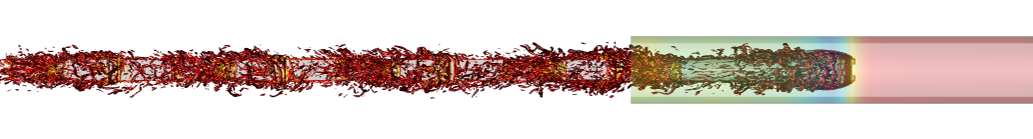}
    \caption*{\text{\small (a)}}
    \end{subfigure}
    \centering   
    \begin{subfigure}[t]{0.9\textwidth}
    \centering
    \includegraphics[width=\textwidth]{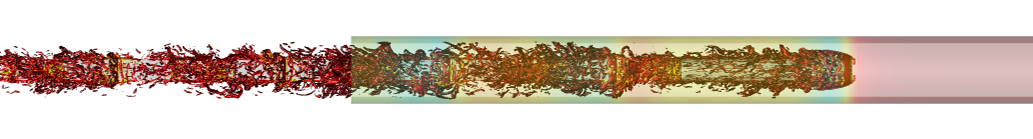}
    \caption*{\text{\small (b)}}
    \end{subfigure}
     \centering   
    \begin{subfigure}[t]{0.9\textwidth}
    \centering
    \includegraphics[width=\textwidth]{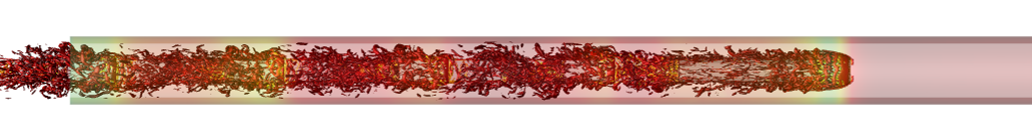}
    \caption*{\text{\small (c)}}
    \end{subfigure}
`    \caption{Velocity contour at plane Y = 0 and distribution of surface wall shear stress at (a) t = 0.03 s (b) t = 0.07 s (c) t = 0.11 s.}
    \label{fig:lambda2}
\end{figure}

Now, due to the restricted area inside the tunnel, the flow speed in the region between the tunnel and train surface accelerates when the train enters. The increasing air velocity and the constraint on flow motion caused by the tunnel walls trigger a higher generation of turbulence. To better visualise the turbulent structures forming around the train, the $\lambda_2$ criterion is adopted to identify the vortices, as shown in Figure~\ref{fig:lambda2}, and the tunnel is coloured by the pressure field. It can be seen that pairs of vortices form at the head and rear of the containers, shedding backwards and influencing the flow field of the downstream containers. Comparing the moments before and after the train enters the tunnel, a noticeable increase in turbulence around the train body is observed once the train is inside the tunnel. Additionally, unlike passenger trains, there is a pronounced 3D pressure phenomenon along the freight train body \citep{soper2014experimental}. The pressure is positive in front of the locomotive/containers, negative at the head of the locomotive/container due to the flow acceleration around the separation bubble, and then gradually decreases along the locomotive/container body. This phenomenon creates a distinctive pressure distribution pattern along the freight train body, which will be quantitatively analysed in more detail in Section~\ref{chap: parameterisationStudy_loadingconfiguration}.

\subsubsection{Validation of the modified 1D code} \label{Chap:Loading_configuration_modification}
\begin{figure}[H]
    \centering   
    \begin{subfigure}[t]{0.49\textwidth}
    \centering
    \includegraphics[width=\textwidth]{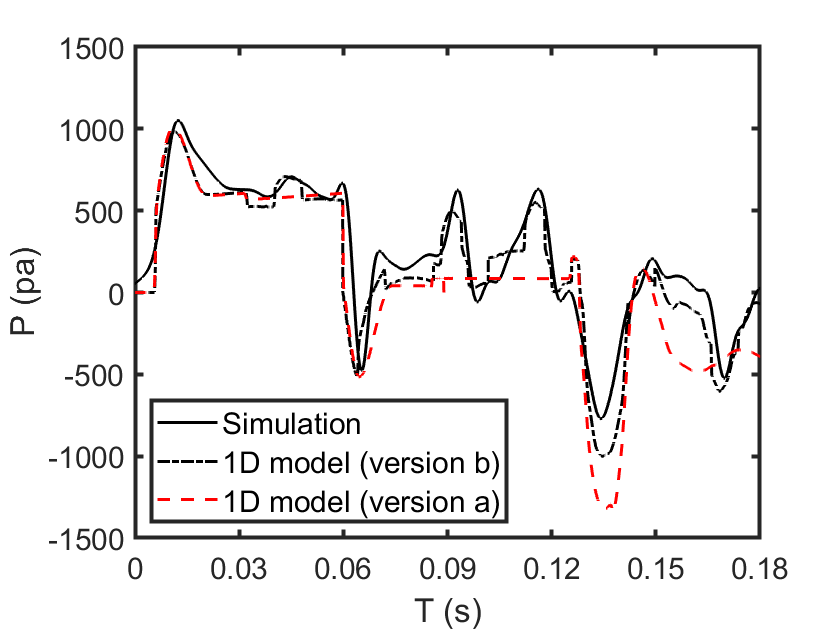}
    \caption*{\text{\small (a)}}
    \end{subfigure}
    \begin{subfigure}[t]{0.49\textwidth}
    \centering
    \includegraphics[width=\textwidth]{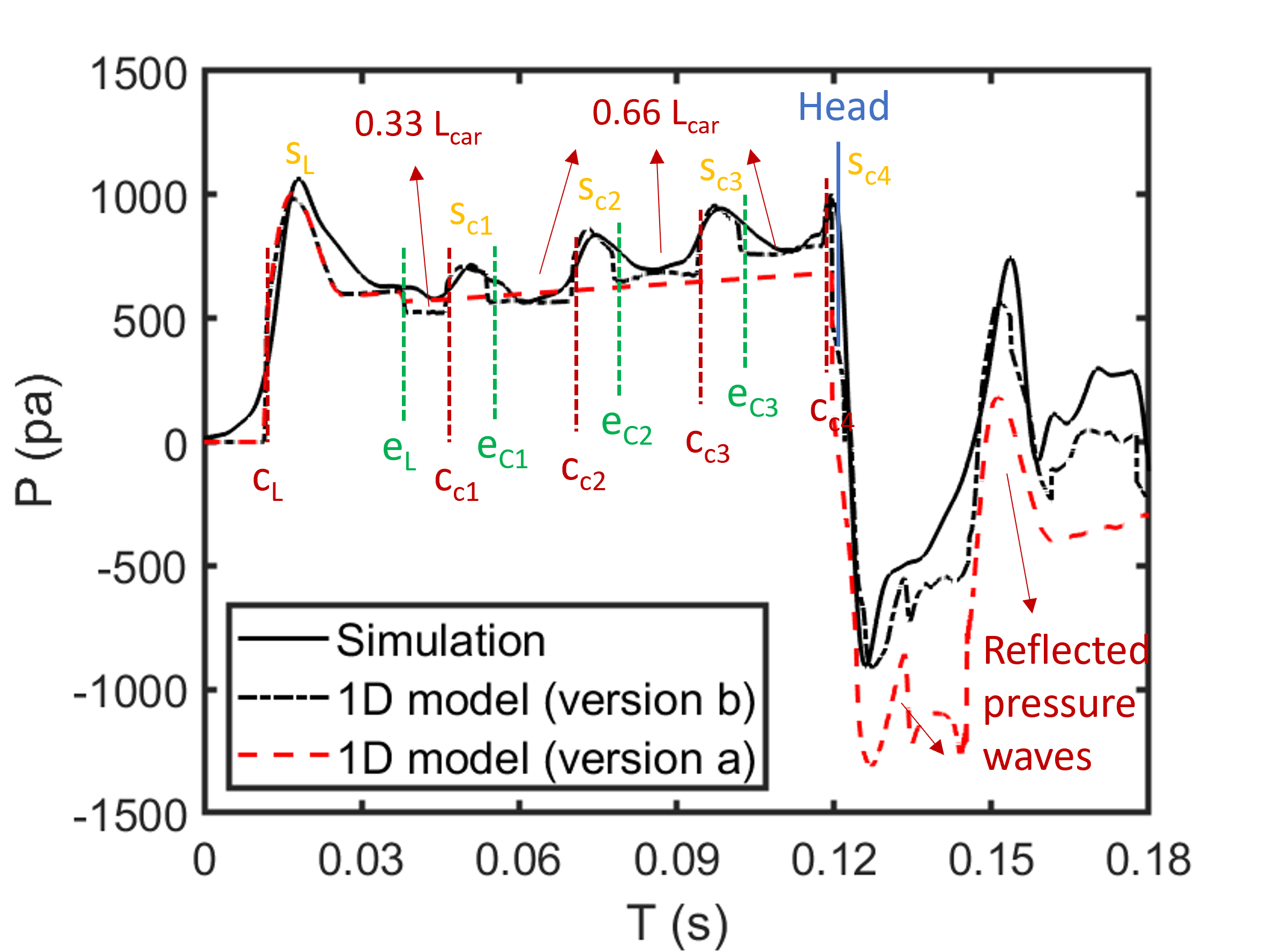}
    \caption*{\text{\small (b)}}
    \end{subfigure}
    \begin{subfigure}[t]{0.49\textwidth}
    \centering
    \includegraphics[width=\textwidth]{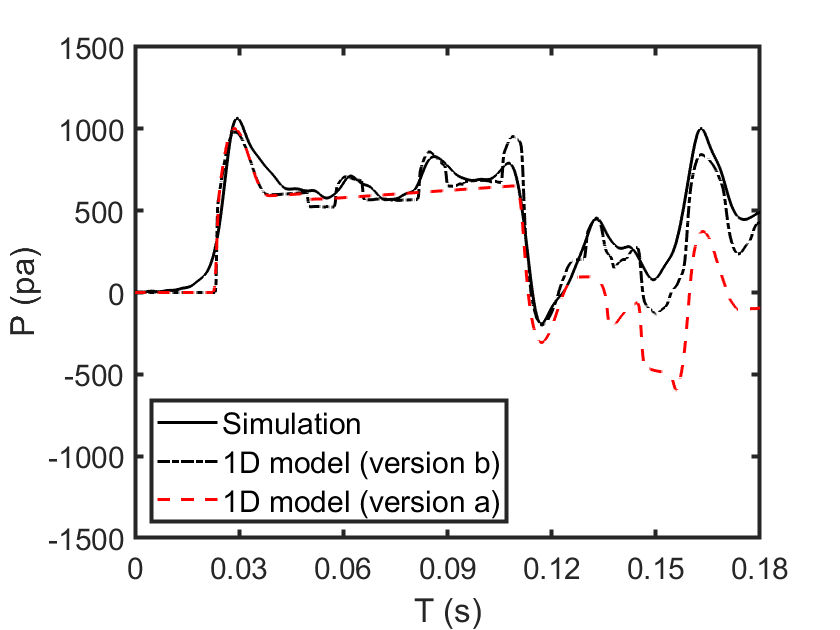}
    \caption*{\text{\small (c)}}
    \end{subfigure}
    \begin{subfigure}[t]{0.49\textwidth}
    \centering
    \includegraphics[width=\textwidth]{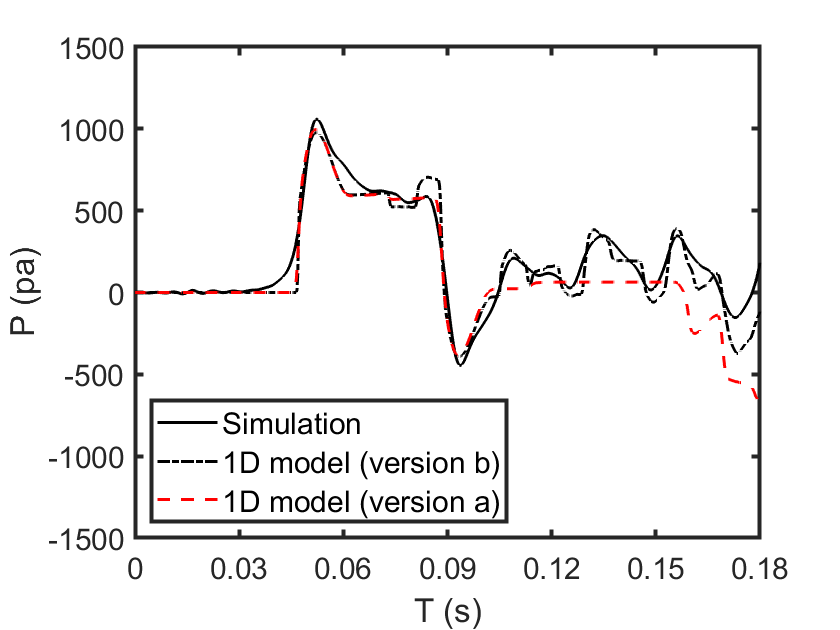}
    \caption*{\text{\small (d)}}
    \end{subfigure}
    \caption{\textcolor{black}{Comparison of pressure history curves obtained from 1D results (version a and b: without and with considering the influence of partially loading configuration) and LES simulation: (a) 2 m (b) 4 m (c) 8 m (d) 16 m.}}
    \label{fig:Loading_configuration_validation}
\end{figure}

The flow field information is used to determine the parameters for the proposed 1D model. These parameters are then used to generate pressure history curves at 2 m, 4 m, 8 m, and 16 m from the tunnel portal, calculated using the modified 1D code. These curves are compared with LES results, as shown in Figure~\ref{fig:Loading_configuration_validation}. The following sections begin by outlining the derivation of the parameters used in the 1D model. In this model, the geometries are characterised by the length, perimeter, effective cross-sectional area, and frictional coefficient of the freight train (locomotive, loaded, and unloaded sections of the container) and the tunnel. The parameters for the locomotive (Class 66) and the tunnel, with a cross-sectional area of 45 $m^2$, remain the same as those presented in \cite{liu2023study}. Notably, from the depiction of the container, it can be seen that the container cross-sectional area (without considering the separation bubble region) is marginally smaller than that of the locomotive, estimated to be approximately 9.8 $m^2$ in the code. 

The frictional coefficient of the freight train body affects the pressure rise, $\Delta p_{fr}$, caused by frictional effects as the main body of the locomotive and containers enter the tunnel. The wall shear stress, $\tau$, is closely related to the frictional coefficient $f$ as expressed by the relation $\tau \approx \frac{1}{2} f \rho u|u|$. Therefore, the distribution of $\tau$ along different parts of the train is used as a reference to determine the frictional coefficient applied in the 1D model. Considering the recirculation "wake region" within the gap, where the large region of negative wall shear stress is observed, the pressure rise stemming from the train body entrance into the tunnel is diminished. Consequently, this reduces the effective frictional coefficient of the entire train. Moreover, the perimeter of the unloaded section is smaller in comparison to the loaded portion of the train, and both these parameters play a role in the computation of the frictional force. For the reference case, where the loaded and unloaded sections of the train are nearly evenly distributed and the respective total lengths are relatively short (within 1 $L_{car}$), both influences are approximated by decreasing the effective frictional coefficient to 0.005 in this study. It is important to note that in future iterations of the model, the influence of these two parameters could be separately simulated and examined in more detail to develop a more refined understanding. This approach not only provides a simplified initial representation, but also offers the potential for more nuanced investigations as the model evolves.

Building upon the discussions in the previous sections, it becomes apparent that the length of the gap significantly influences the internal flow structure within the gap itself, as well as the separation bubble generated by the airflow over the container behind it. For a 33$\%$ partially loaded freight train, the gap between the locomotive and the first container is around 6 m (33$\%$ of a carriage) and the gap between the containers for the following containers are approximately 12 m (66$\%$ of a carriage length). For the sake of clarity, the length of a whole carriage will be denoted as $L_{\text{car}}$ in the subsequent discussions. Consequently, it follows that the parameters allocated for the first gap (33$\%$ $L_{\text{car}}$) within the 1D code differ from those employed for the subsequent gaps (66$\%$ $L_{\text{car}}$). In the case of the first gap, the gap length is shorter than the reattachment length. Subsequently, the following container restricts the wake development, causing a significant portion of the airflow to be trapped within the gap. This distinction from the 12 m gap results in a comparatively larger effective cross-sectional area in the 6 m gap configuration. Consequently, this configuration exhibits reduced pressure loss, particularly in tandem with both the leading and trailing parts of the train. Moreover, the transition of air from the gap to the subsequent container occurs more smoothly in the 6 m gap configuration. 

\begin{figure}[htbp]
    \centering
    \includegraphics[width=16cm]{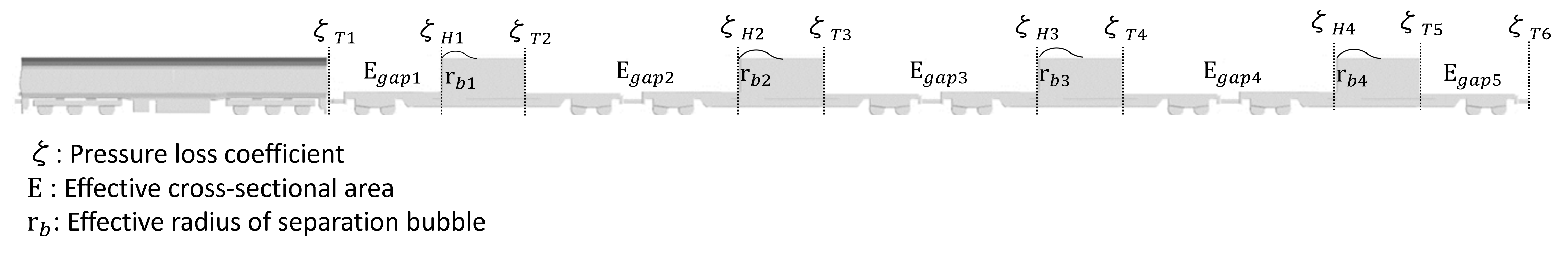}
    \caption{Illustrative figure of the pre-determined parameters used in the 1D code for the 33$\%$ loaded freight train.}
    \label{fig:reference case}
\end{figure}

This smoother flow transition leads to a smaller separation bubble size forming at the front of the container located after the 6 m gap. It is important to note that using LES simulations, the transient and highly turbulent nature of the flow field makes capturing the precise separation bubble shape challenging. Given that the relationship between separation bubble size and the pressure rise induced by pressure waves during train tunnel entry has been previously established and analysed in \cite{liu2023study}, a best-fit method \citep{vardy1999estimation} is employed to deduce the parameters governing the separation bubble sizes for the containers after the 33$\%$ $L_{\text{car}}$ and 66$\%$ $L_{\text{car}}$ gaps. In summary, the predetermined parameters adopted in the 1D code for the locomotive with four $33 \%$ freight cars are illustrated in  Figure~\ref{fig:reference case} and listed in Table~\ref{tab:reference case}. \textcolor{black}{These parameters are subsequently implemented into the 1D code. After modifying the mesh system and boundary conditions, the updated 1D model is rerun to compute the pressure variations induced by a $33\%$ loaded freight train carrying four containers. The resulting pressure history curves inside the tunnel, calculated using the modified 1D code (version b: dashed black line), are compared with the LES results, as shown in Figure~\ref{fig:Loading_configuration_validation}. To highlight the impact of incorporating partial loading configurations into the 1D model, results from an earlier version of the code (version a: dashed red line) are also included for comparison.} It can be observed that the pressure change caused by the compression wave generated by the first container ($c_{c1}$) is smaller than those generated by the subsequent containers ($c_{c2}$ to $c_{c4}$). Moreover, due to the larger separation bubble formed at the front of the containers behind longer unloaded gaps ($s_{c2}$ to $s_{c4}$ as marked in Figure~\ref{fig:Loading_configuration_validation}(b)) compared to $s_{c1}$, these larger separation bubble regions also have a greater influence on the magnitude of the pressure wave pattern. Similarly, the pressure drop caused by the expansion wave from the locomotive's tail ($e_{L}$) is smaller than those generated by the tails of the following containers ($e_{c1}$ to $e_{c3}$). This difference arises because the gap between the locomotive and the first container (33$\%$ $L_{\text{car}}$) is shorter than the gaps between the unloaded containers (66$\%$ $L_{\text{car}}$). The wake flow from the forward container cannot fully develop and is confined within the shorter gap. This leads to a larger effective cross-sectional area in the gap and a reduced area difference between the loaded and unloaded regions, which in turn weakens the pressure waves. Overall, the step-wise pressure rise caused by the sub-pressure waves when the partially loaded containers enter into the tunnel, along with the variations in the magnitudes of these pressure waves, can be seen to be  accurately simulated after modifications to the new 1D code (version(b)). Moreover, the propagation and magnitude of these sub-pressure waves align well with the LES results. \textcolor{black}{In comparison, the previous version (a) exhibits a clear limitation: it fails to capture the sub-pressure waves generated by the partially loaded containers and the associated reflected waves. As a result, it is unable to reproduce the overall pressure wave pattern and the magnitude of the pressure changes induced by the container entries, leading to significant discrepancies from the LES results.} It is worth noting that the parameters obtained from the best-fit method are utilised for validating the feasibility of the proposed algorithms in Section~\ref{sec:1D_BC}, while more comprehensive analysis will be conducted in subsequent sections through the examination of the mean flow fields from LES simulations of freight trains featuring diverse loading configurations.

\begin{table}[htbp]
    \centering
    \begin{tabular}{|c|c|}
        \hline
        \textbf{Parameter} & \textbf{Value} \\
        \hline
        Length (m) & $L_{con} = 6.096$, $L_{gap1} = 6.096$, $L_{gap2} \sim L_{gap4} = 12.192$ \\
        \hline
        Effective Area (m$^2$) & $E_{con} = 9.8$, $E_{gap1} = 0.84 \times E_{con}$, $E_{gap2} \sim E_{gap4} = 0.7 \times E_{con}$ \\
        \hline
        Head loss $\zeta_H$ & $\zeta_{H1} = 0.06$, $\zeta_{H2} \sim \zeta_{H3} = 0.08$ \\
        \hline
        Tail loss $\zeta_T$ & $\zeta_{Tn} = \left(\frac{E_{con} - E_{gapn}}{E_{tu} - E_{con}}\right)^2$ \\
        \hline
        Effective speration bubble radius (m) & $r_{b1} = 0.05$, $r_{b2} \sim r_{b4} = 0.085$ \\
        \hline
    \end{tabular}
    \caption{Parameters of the containers used in the modified 1D code (Full scale).}
    \label{tab:reference case}
\end{table}

\subsection{Influence of different loading configurations} \label{chap: parameterisationStudy_loadingconfiguration}
\subsubsection{Streamline and velocity field}
\begin{figure}[H]
    \centering   
    \begin{subfigure}[t]{0.35\textwidth}
    \centering
    \includegraphics[width=\textwidth]{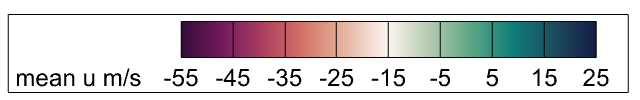}
    \end{subfigure}
    \centering   
    \begin{subfigure}[t]{0.35\textwidth}
    \centering
    \includegraphics[width=\textwidth]{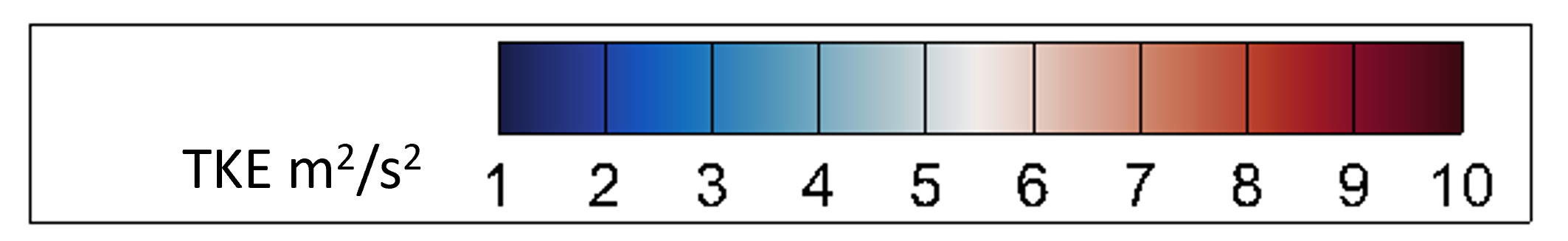}
    \end{subfigure}
    \centering   
    \begin{subfigure}[t]{0.49\textwidth}
    \centering
    \includegraphics[width=\textwidth]{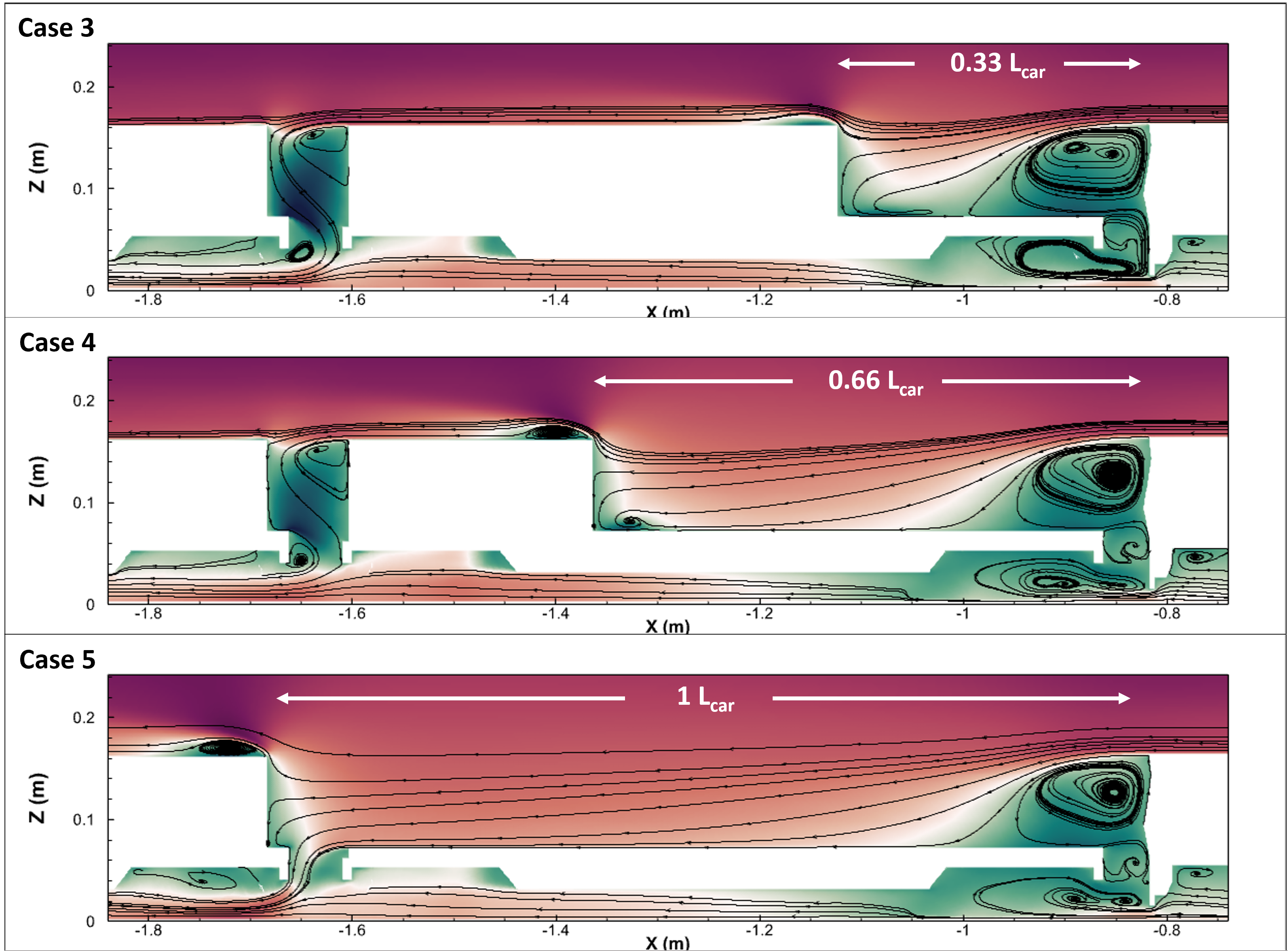}
    \caption*{\text{\small (a)}}
    \end{subfigure}
    \begin{subfigure}[t]{0.49\textwidth}
    \centering
    \includegraphics[width=\textwidth]{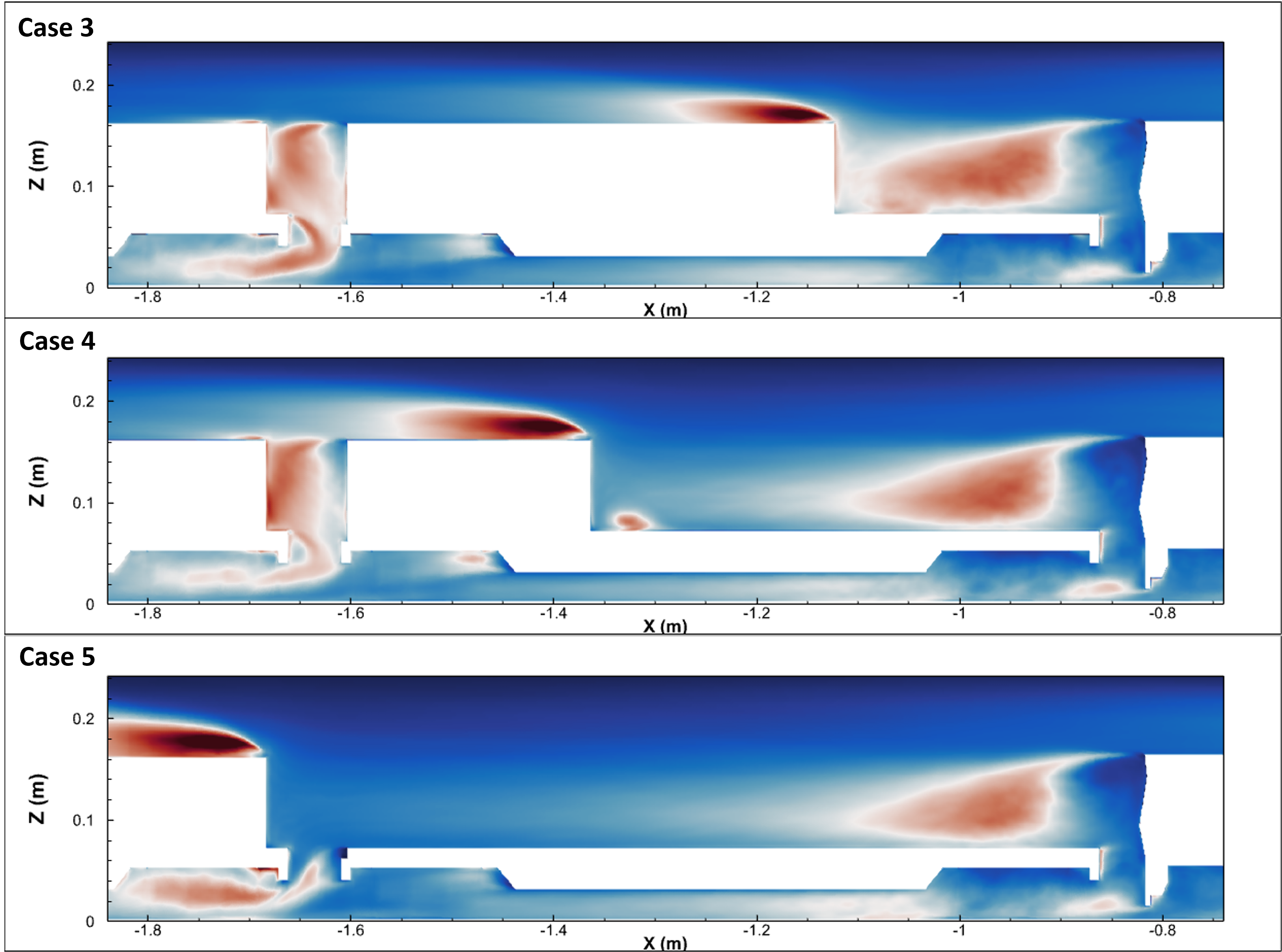}
    \caption*{\text{\small (b)}}
    \end{subfigure}
    \caption{Velocity field of the first wagon at contour Y=0 m (a) mean $u$ and streamline (b) TKE.}
    \label{fig:u_container1y1}
\end{figure}

A parameterisation study is carried out by simulating different container loading patterns for the intermodal freight train passage through the tunnel. For this study three loading efficiencies were adopted, namely 66$\%$/33$\%$/0$\%$ loaded. To visualise the wake vortex formation as the air flows over the leading carriage, Figure~\ref{fig:u_container1y1}  displays the mean velocity distribution and streamlines of container 1 for each loading configuration separately at Y=0 m and Z=0.09 m. The three sub-figures in each of these figures correspond to three loading configurations, as outlined above, resulting in gap distances between the locomotive and the loaded part equivalent to 0.33 $L_{\text{car}}$, 0.66 $L_{\text{car}}$, and 1 $L_{\text{car}}$ respectively. The subsequent figures in this section maintain the same order. The left side of the contours is coloured by $\Bar{u}$ (mean stream-wise velocity), while the right side is coloured by turbulent kinetic energy (TKE). TKE is calculated as the sum of root-mean-square velocity fluctuations in three directions (as defined in Equation \ref{eq:TKE}). This TKE parameter serves as a measure of the turbulence generation and overall turbulence intensity surrounding the train.

\begin{equation}
    {TKE}=0.5\times({u^\prime}^2+{v^\prime}^2+{w^\prime}^2)
\label{eq:TKE}
\end{equation}

A consistent observation across all three loading configurations is the separation of the boundary layer at the end of the locomotive, resulting in the formation of large vortices. The length of the recirculation wake region is approximately 0.25 times the length of the container (0.25 $L_{\text{car}}$), which is shorter than the smallest gap size (0.33 $L_{\text{car}}$). In the cases with gap lengths of 0.33 and 0.66 $L_{\text{car}}$, the trailing container is positioned at the wake reattachment zone. Some of the air downwash from the leading container is obstructed by the trailing container, leading to the formation of a backflow region in front of it. For the 0.33 $L_{\text{car}}$ gap case, where the gap is quite short, this backflow region connects with and influences the wake recirculation region. The contours in the figures illustrate that the length of the wake vortex is larger, with higher TKE compared with the other two cases. 

Additionally, since the container itself constitutes a bluff body with a sharp leading edge, separation bubbles form at the top and sides of the container. The size of the separation bubble around the head of the trailing container significantly increases with a longer gap length. This observation aligns with the earlier discussion, where it was established that the oncoming flow velocity increases as the gap length becomes longer. The heights of the separation bubbles at the top and sides of the container are summarised in Table~\ref{tab:separationbubble_container}. The position of the slices used to measure $H_{R_c}/W_{C}$ and $H_{S_c}/W_{C}$ are at Y=0 m and Z=0.12m, which are approximately at the middle of the container width and height. These heights are non-dimensionalised using the width of the container, denoted as $W_c$. These values will play a crucial role in calculating the effective blockage area at the containers, to be fed back into the 1D model. The separation bubble model for containers and the parameterisation equation correlating $H_{R_c}$, $H_{S_c}$, and the gap length are further explored in section~\ref{chap:parameterisation_equation}.

\begin{table}[H]
  \footnotesize    
  \begin{center}
  \def~{\hphantom{0}}
  \begin{tabular}{lccc}
      \hline
      Gap Length & 0.33 $L_{\text{car}}$ & 0.66 $L_{\text{car}}$ & 1 $L_{\text{car}}$  \\
      \hline
      \(H_{R_c}/W_{C}\) & 0.082 & 0.1327 & 0.158 \\
      \(H_{S_c}/W_{C}\) & 0.071 & 0.1122 & 0.154\\
      \hline
  \end{tabular}
    \caption{The height of the separation bubble at the top and \textcolor{black}{sides} of the container.}
  \label{tab:separationbubble_container}
  \end{center}
\end{table}

\subsubsection{Significant zones around partially loaded freight train}

\begin{figure}[H]
    \centering
    \begin{subfigure}[t]{0.49\textwidth}
    \centering
    \includegraphics[width=\textwidth]{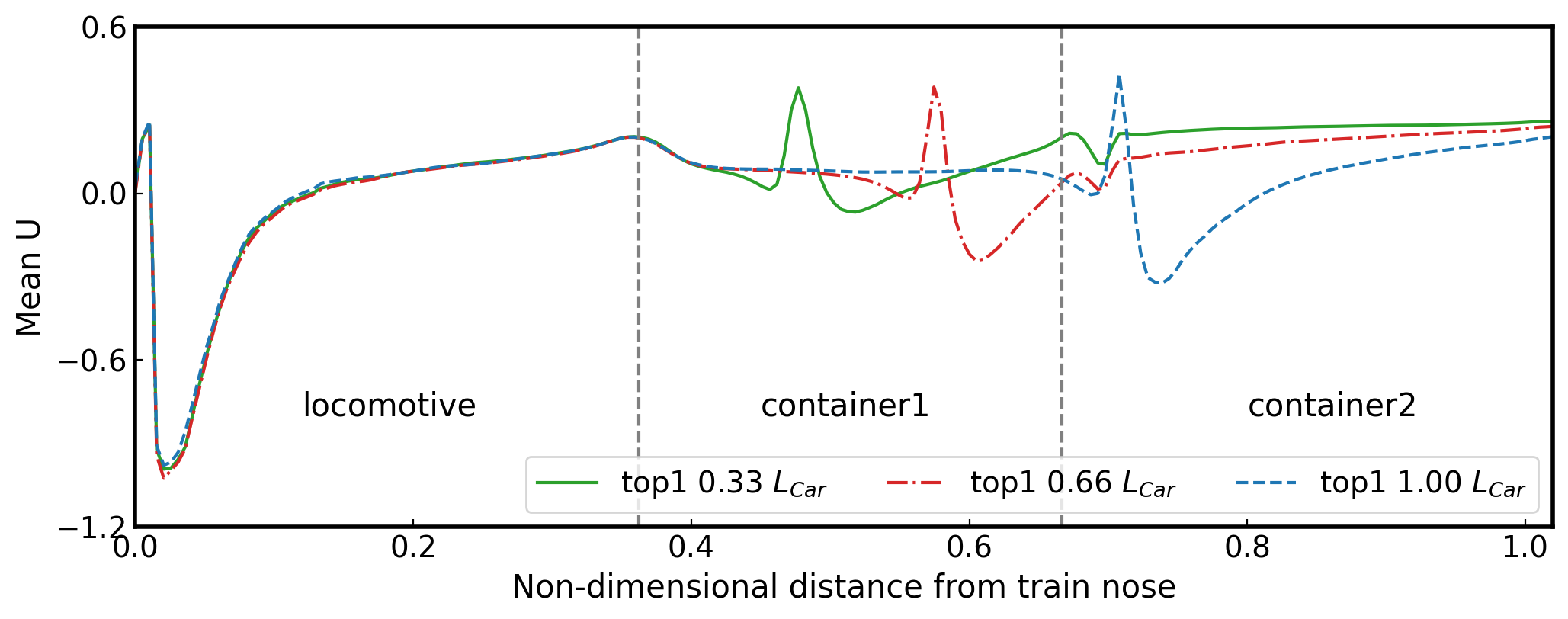}
    \caption*{\text{\small (a)}}
    \end{subfigure}
    \begin{subfigure}[t]{0.49\textwidth}
    \centering
    \includegraphics[width=\textwidth]{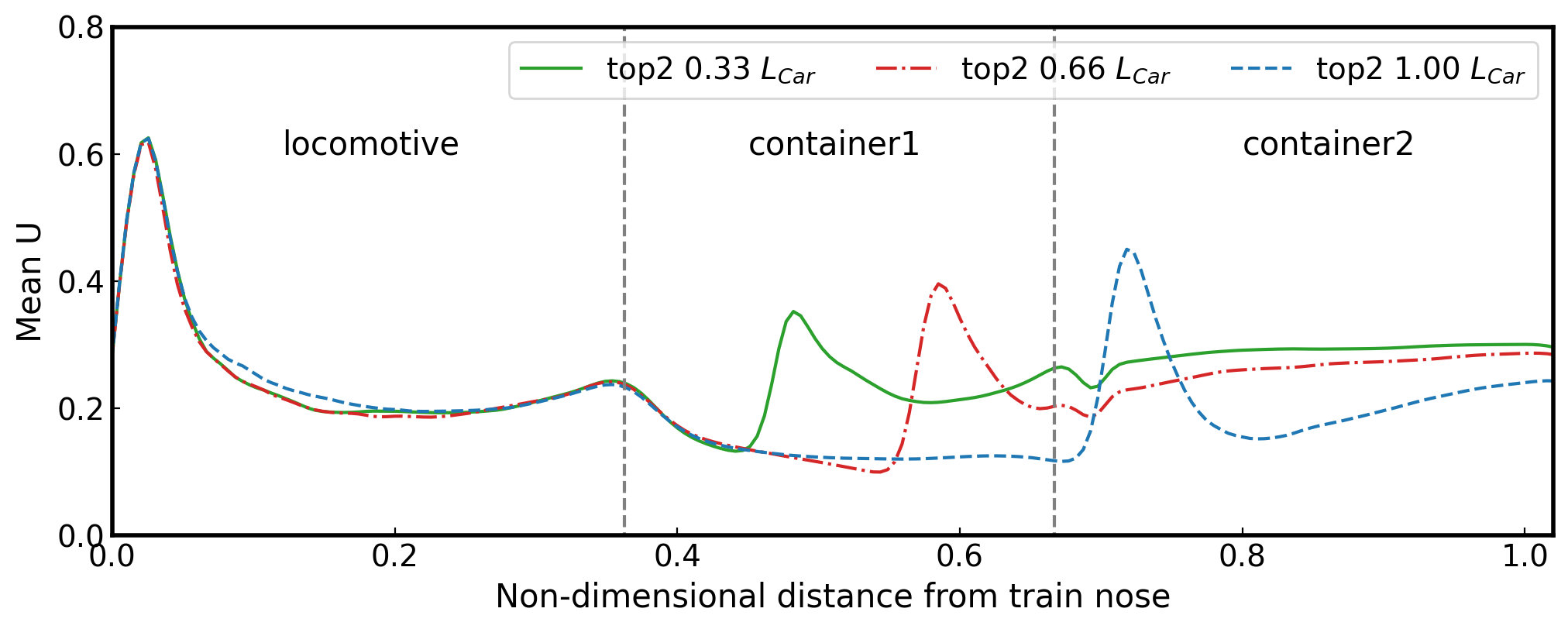}
    \caption*{\text{\small (b)}}
    \end{subfigure}
    \begin{subfigure}[t]{0.49\textwidth}
    \centering
    \includegraphics[width=\textwidth]{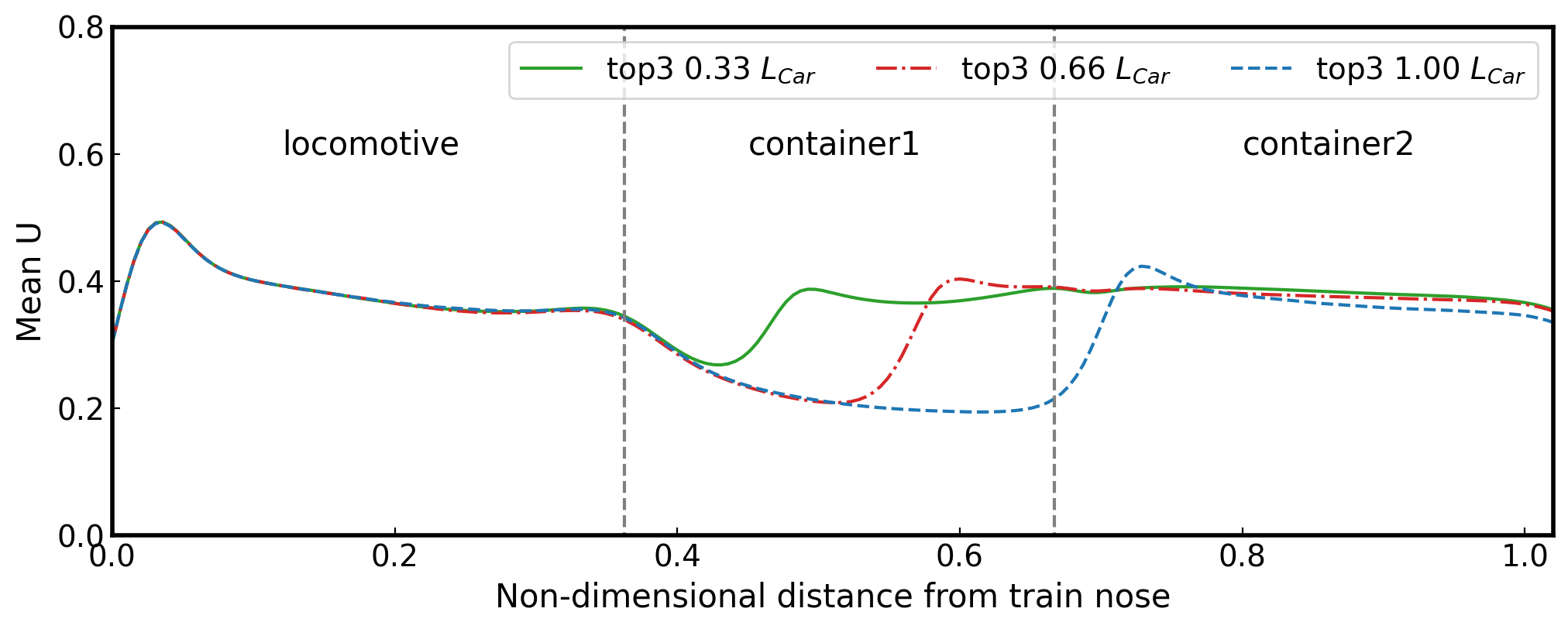}
    \caption*{\text{\small (c)}}
    \end{subfigure}
    \begin{subfigure}[t]{0.49\textwidth}
    \centering
    \includegraphics[width=\textwidth]{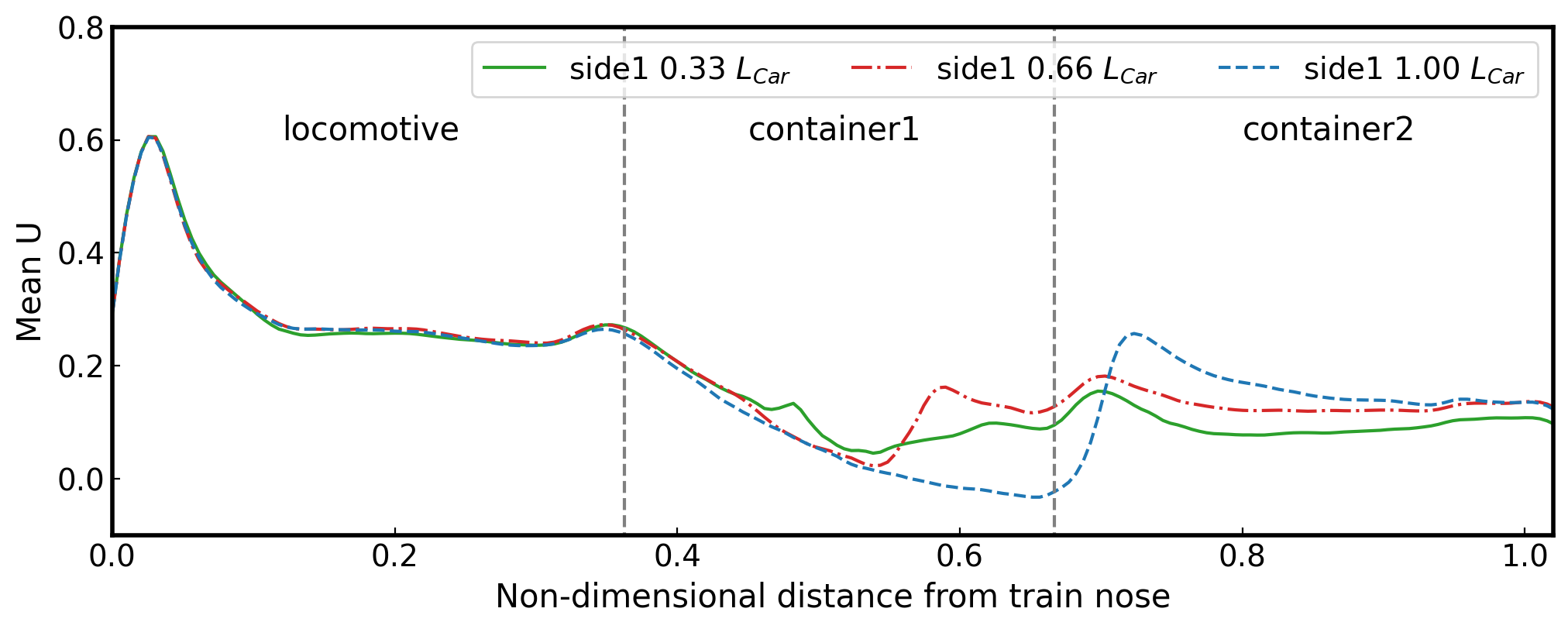}
    \caption*{\text{\small (d)}}
    \end{subfigure}
    \begin{subfigure}[t]{0.49\textwidth}
    \centering
    \includegraphics[width=\textwidth]{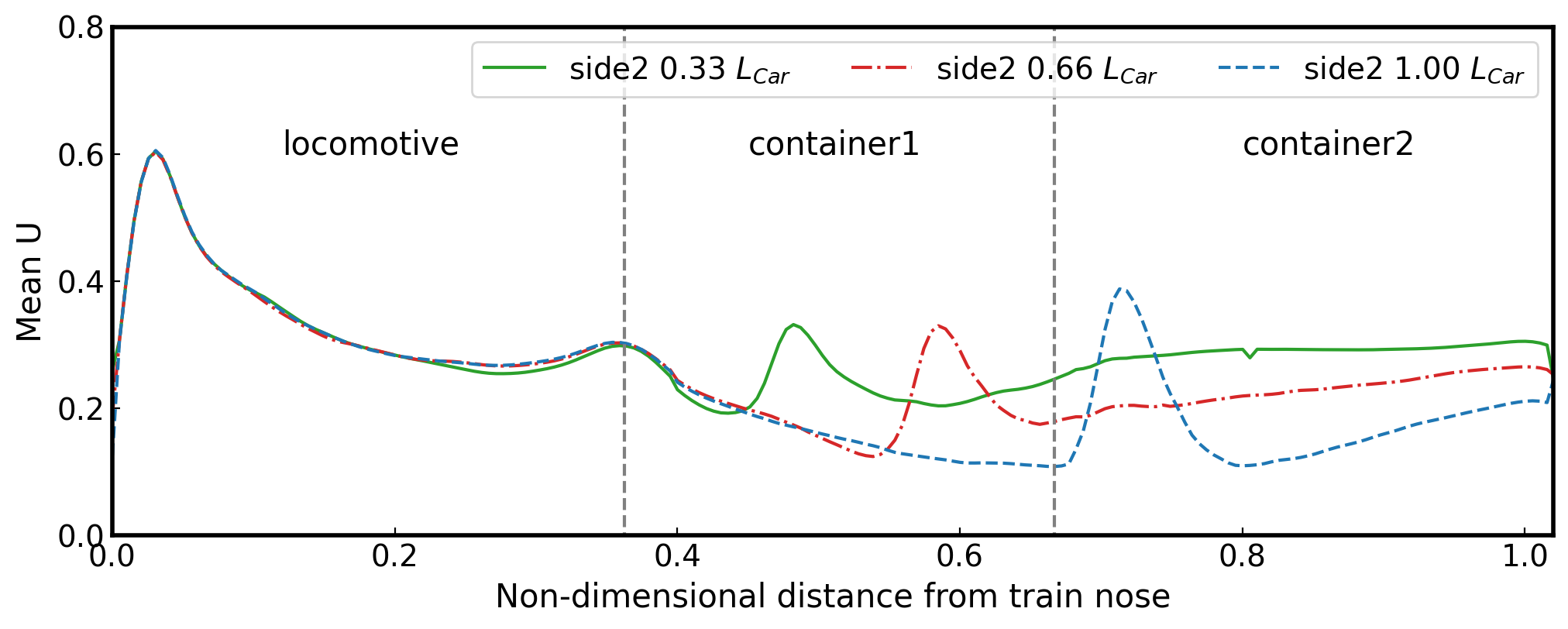}
    \caption*{\text{\small (e)}}
    \end{subfigure}
        \begin{subfigure}[t]{0.49\textwidth}
    \centering
    \includegraphics[width=\textwidth]{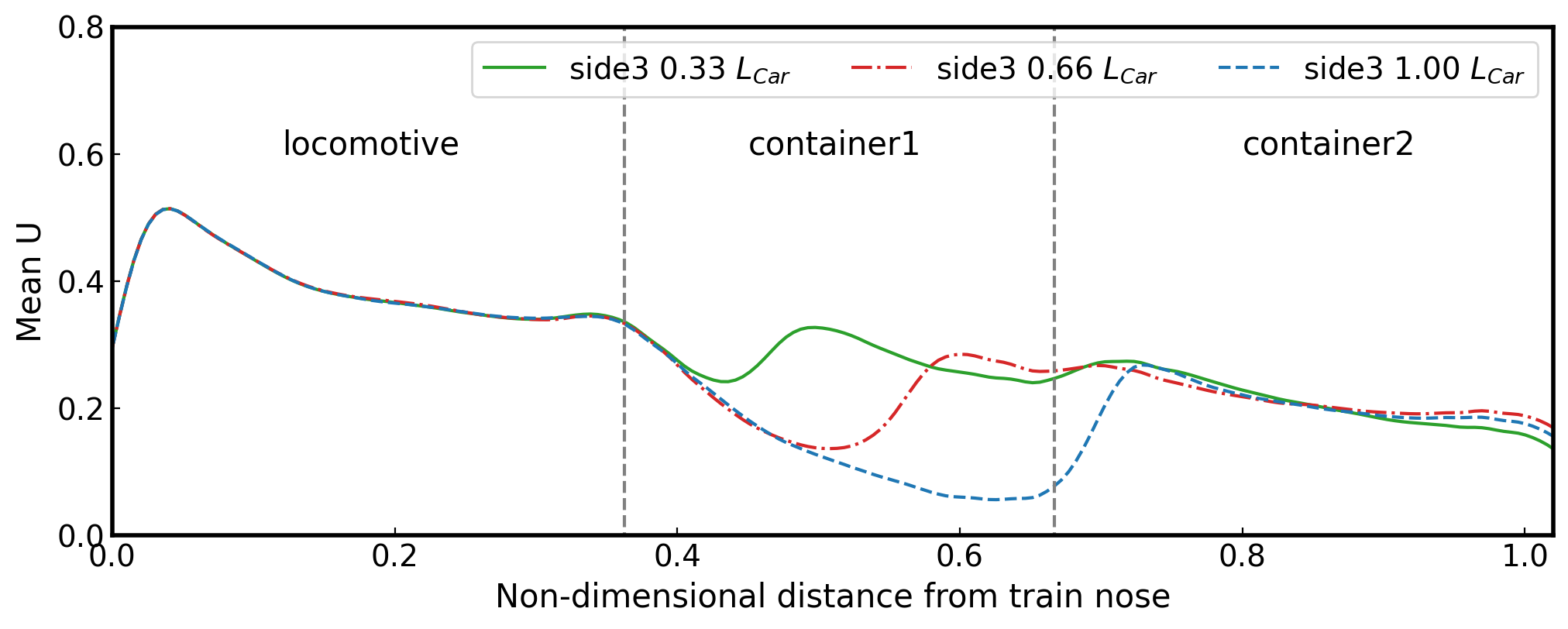}
    \caption*{\text{\small (f)}}
    \end{subfigure}
    \caption{Distribution of mean $\Bar{U}$ of three loading configurations at different lines positioned along the train (a) top1 (b) top2 (c) top3 (d) side1 (e) side2 (f) side3.}
    \label{fig:meanu_lines_streamwise}
\end{figure}

To conduct a quantitative analysis of the velocity magnitude across various loading configurations, the variation of $\Bar{U}$ is plotted along distinct lines positioned along the entire length of the freight train (Figure~\ref{fig:meanu_lines_streamwise}). Here, $\Bar{U}$ represents the mean streamwise velocity relative to the ground and is non-dimensionalised by the train operational speed. The positive direction of $\Bar{U}$ is denoted from the head to tail of the train. Additionally, the length of the lines are non-dimensionalised by the total length of the freight train. From the figures, it is evident that the velocity characteristics in freight trains with partially loaded containers differ from those observed in passenger trains, due to the discontinuity of the train body caused by the varying container loading conditions. 

The lines labelled as top1/2/3 are positioned at heights of 0.02/0.04/0.08 m above the train roof, respectively. Additionally, the lines labelled sides1/2/3 are situated 0.04 m away from the train side, with varying heights of 0.06/0.09/0.12 m, respectively. Considering Z=0.6 m corresponds to a full scale of 1.5 m, which is crucial for human safety, an additional measuring position (side4) is introduced at this height, located 0.08 m away from the train side. The positions of these lines relative to the locomotive and container are clearly depicted in Figure~\ref{fig:line_positions}. To facilitate a comprehensive investigation of the velocity characteristics along train carriages with different loading configurations, the three carriages are divided and labelled in each sub-figure. The measuring lines, except top1, are positioned beyond the recirculation zones in each separation bubble region, and share similar velocity distribution pattern among different lines. In summary, referenced by the method to identify the flow regions of passenger trains \citep{bell2020boundary,soper2014experimental,baker2014full}, the velocity distribution around an intermodal container freight train can be characterised by a series of zones as follows:

\begin{enumerate}
    \item Starting from the blunt head of the locomotive, the first separation bubble zone ($Zone_{S1}$) is formed, where the magnitude of the velocity is primarily influenced by the relative position of the measuring points with respect to the separation bubble. A noticeable velocity peak is observed due to flow acceleration around the separation bubble.
    \item Subsequently, a boundary layer zone ($Zone_{B1}$) follows, in which the magnitude of $\Bar{U}$ at measuring points located close to the train body (lines top/side 1 - 2) gradually increases due to the development and thickening of the boundary layer. 
    \item As the airflow reaches the tail of the locomotive, the boundary layer separates, leading to the formation of a quasi-wake region ($Zone_{W1}$) at the gap. The velocity gradually decreases as the distance to the locomotive tail increases.
    \item The airflow then flows over the partially loaded container and generates the second separation bubble zone ($Zone_{S2}$), where a local peak velocity is observed.
    \item Similarly, a second boundary layer zone ($Zone_{B2}$) develops along the body of the fully-loaded containers with $\Bar{U}$ gradually decreasing.
\end{enumerate}

 This pattern of velocity characterisation can be similarly deduced for scenarios involving more gaps and containers in a partially loaded freight train. Furthermore, it can be observed from the figure that there is an obvious 3d effect on $\Bar{U}$ at $Zone_{S1}$ and $Zone_{S2}$. An intriguing observation emerges with regard to the velocity distribution along line top1, which is situated within the separation bubble ($Zone_{S1}$). This line passes through the centre of the separation bubble generated at the front of the locomotive, resulting in a $\Bar{U} \approx -1$ (corresponding to a velocity close to train velocity). Conversely, for the lines positioned outside the separation bubble top2/3, an acceleration is observed as the airflow passes over them. A similar phenomenon can be observed at $Zone_{S2}$ when the air flows over the loaded container behind the initial gap. Regarding the flow characteristics at the partially loaded containers, the line top1 traverses the upper part of the separation bubble generated as the airflow passes over the container gap. As anticipated, the larger gap (1 $L_{car}$) induces a larger separation bubble at the top of the container, leading to the line top1 ( Figure~\ref{fig:meanu_lines_streamwise}(a)) being closer to the vortex centre and the magnitude of $\Bar{U}$ is therefore closer to the train velocity. Conversely, for the other two cases, line top1 is situated closer to the upper edge of the vortex, resulting in $\Bar{U}$ being closer to the upstream flow velocity. The peak value of $\Bar{U}$ of line top1 at the the separation bubble generated by the container changes from around -0.46/-0.21/-0.11  when the gap changes from 1/0.66/0.33 of wagon length. 

\begin{figure}[htbp]
    \centering
    \begin{subfigure}[t]{0.4\textwidth}
    \centering
    \includegraphics[width=\textwidth]{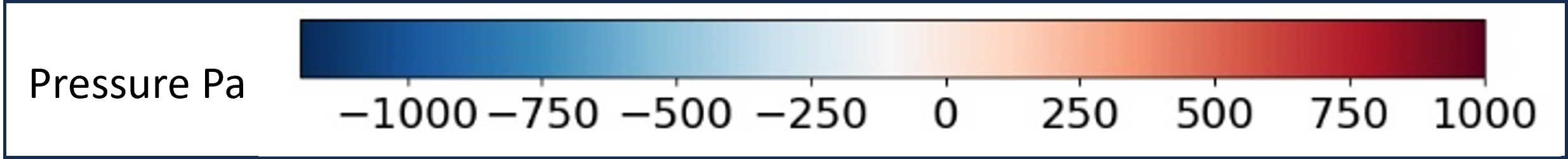}
    \end{subfigure}
    \centering
    \begin{subfigure}[t]{0.6\textwidth}
    \centering
    \includegraphics[width=\textwidth]{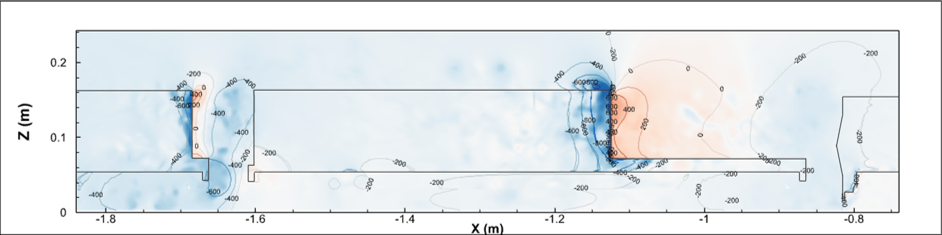}
    \caption*{\text{\small (a)}}
    \end{subfigure}
    \begin{subfigure}[t]{0.6\textwidth}
    \centering
    \includegraphics[width=\textwidth]{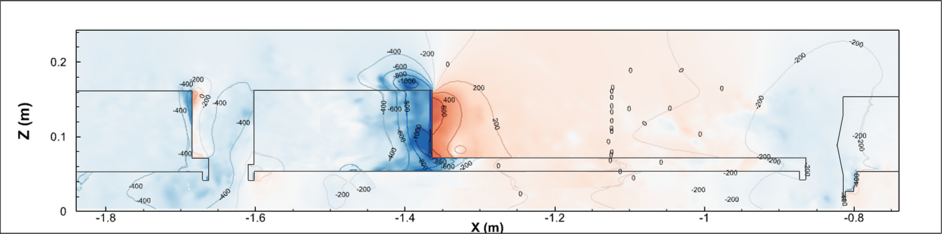}
    \caption*{\text{\small (b)}}
    \end{subfigure}
    \begin{subfigure}[t]{0.6\textwidth}
    \centering
    \includegraphics[width=\textwidth]{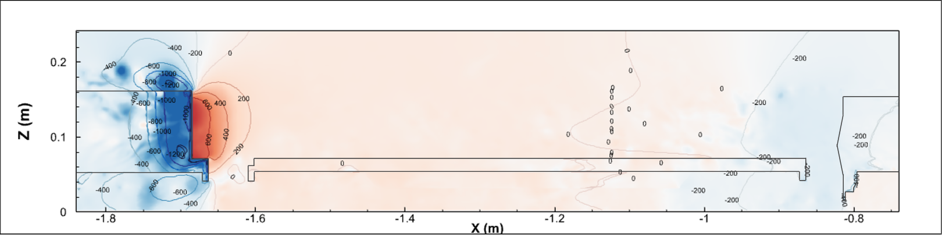}
    \caption*{\text{\small (c)}}
    \end{subfigure}
    \caption{Pressure distribution around the first wagon with gap distances of (a) 0.33 $L_{\text{car}}$ (b) 0.66 $L_{\text{car}}$ and (c) 1 $L_{\text{car}}$.}
    \label{fig:p_container1}
\end{figure}

The pressure distributions at the partially loaded container are depicted in Figure~\ref{fig:p_container1}. The pressure contours are coloured based on instantaneous pressure, and contour lines are drawn using mean pressure data. In all of the partially loaded cases studied, a common phenomenon observed is that of the initial formation of a weak negative pressure zone behind the locomotive, corresponding to the wake recirculation region ($Zone_{W1}$). Following this, a positive pressure region is observed in front of the loaded container. The maximum pressure ahead of the container increases with the gap length. Specifically, it rises from approximately 200 to 400/600 Pa as the gap increases from 0.33 to 0.66/1 $L_{car}$. The presence of a negative pressure region around the container head indicates the presence of a separation bubble, corresponding to the second separation bubble region $Zone_{S2}$. Furthermore, the length and magnitude of the negative pressure region around the head of the loaded container also increase significantly with the extension of the unloaded gap. 

\begin{figure}[H]
    \centering
    \begin{subfigure}[t]{0.49\textwidth}
    \centering
    \includegraphics[width=\textwidth]{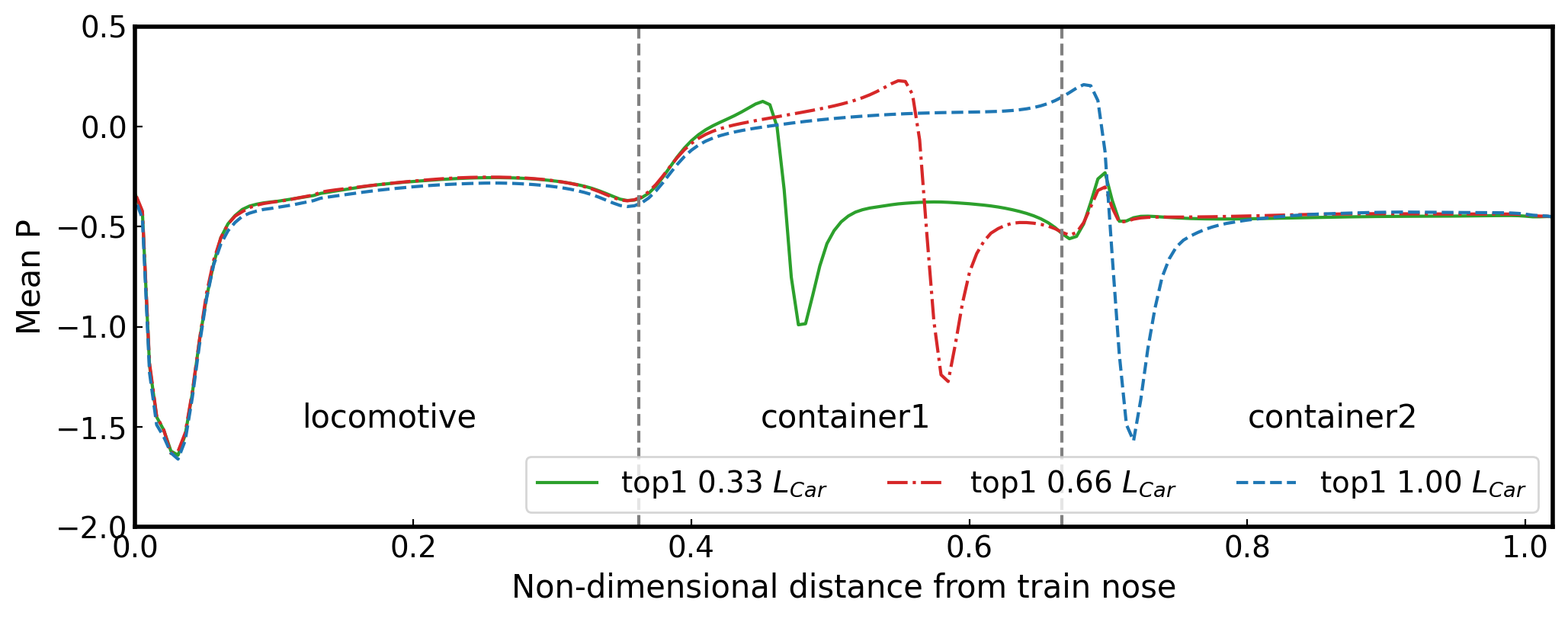}
    \caption*{\text{\small (a)}}
    \end{subfigure}
    \begin{subfigure}[t]{0.49\textwidth}
    \centering
    \includegraphics[width=\textwidth]{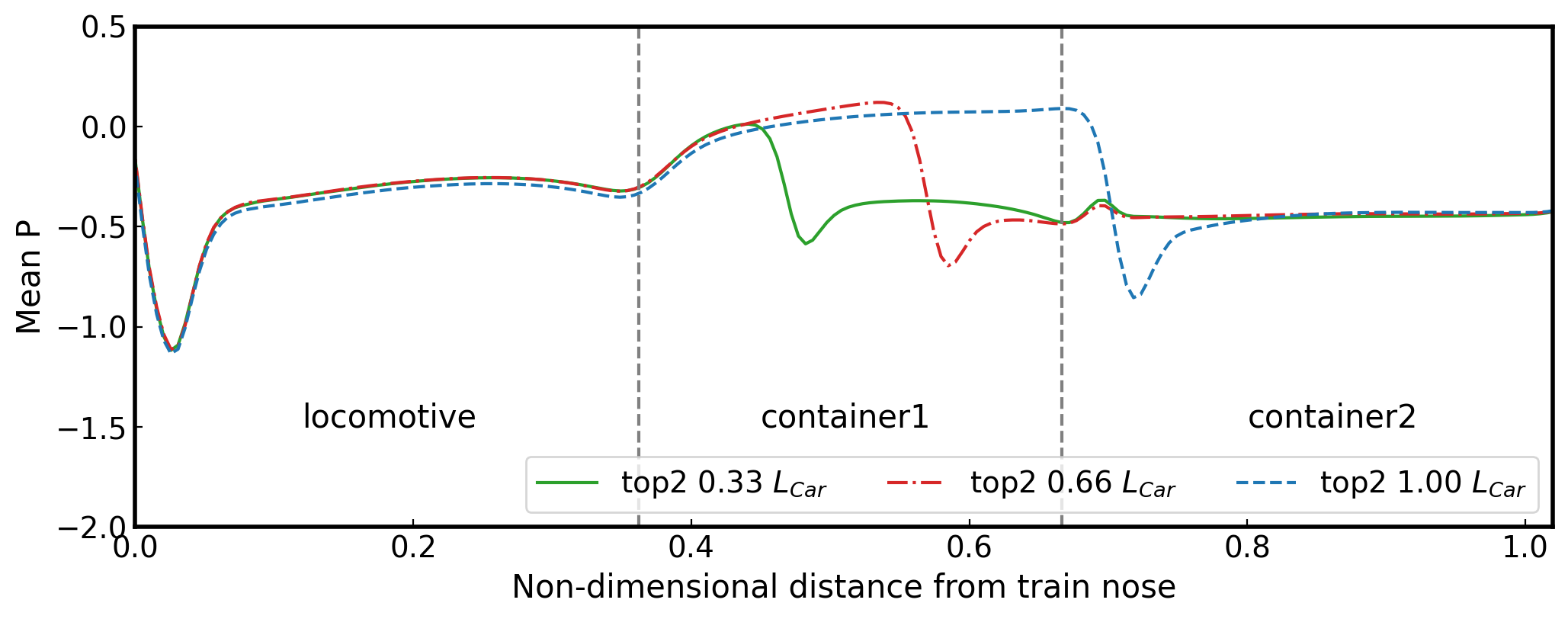}
    \caption*{\text{\small (b)}}
    \end{subfigure}
    \begin{subfigure}[t]{0.49\textwidth}
    \centering
    \includegraphics[width=\textwidth]{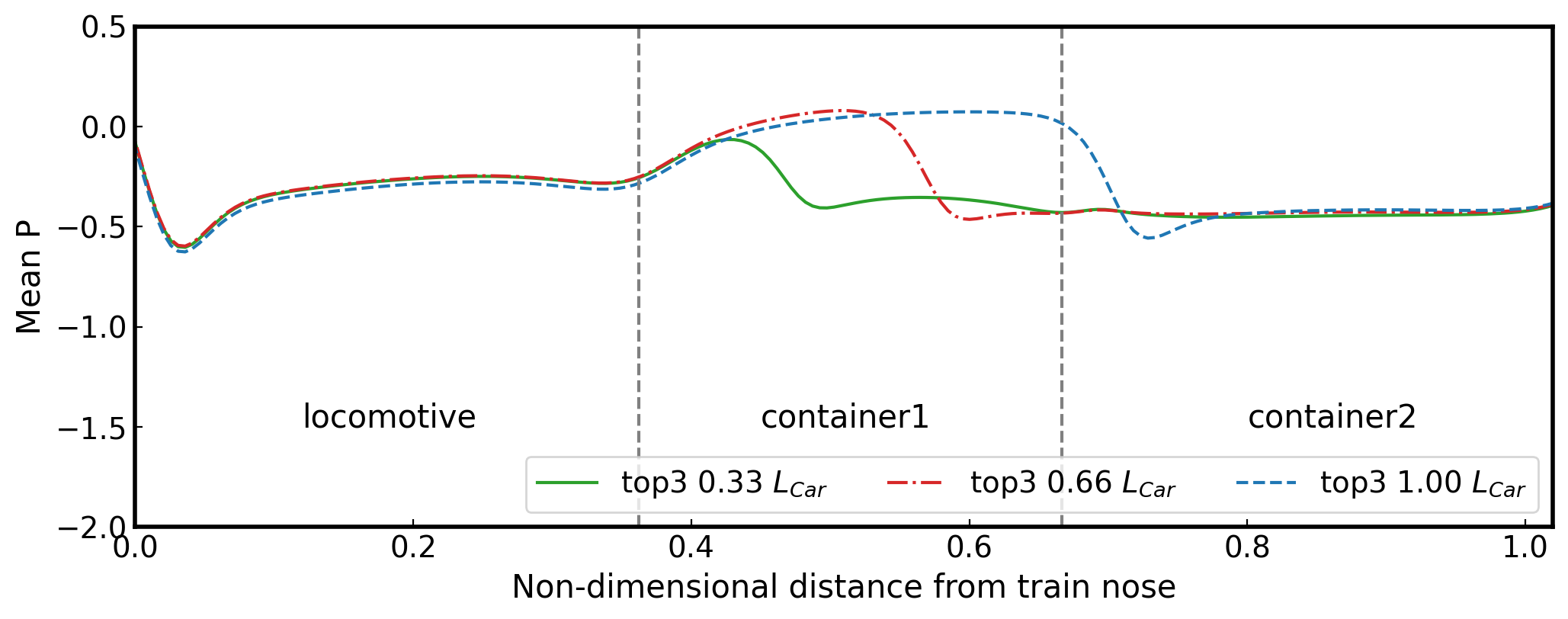}
    \caption*{\text{\small (c)}}
    \end{subfigure}
    \begin{subfigure}[t]{0.49\textwidth}
    \centering
    \includegraphics[width=\textwidth]{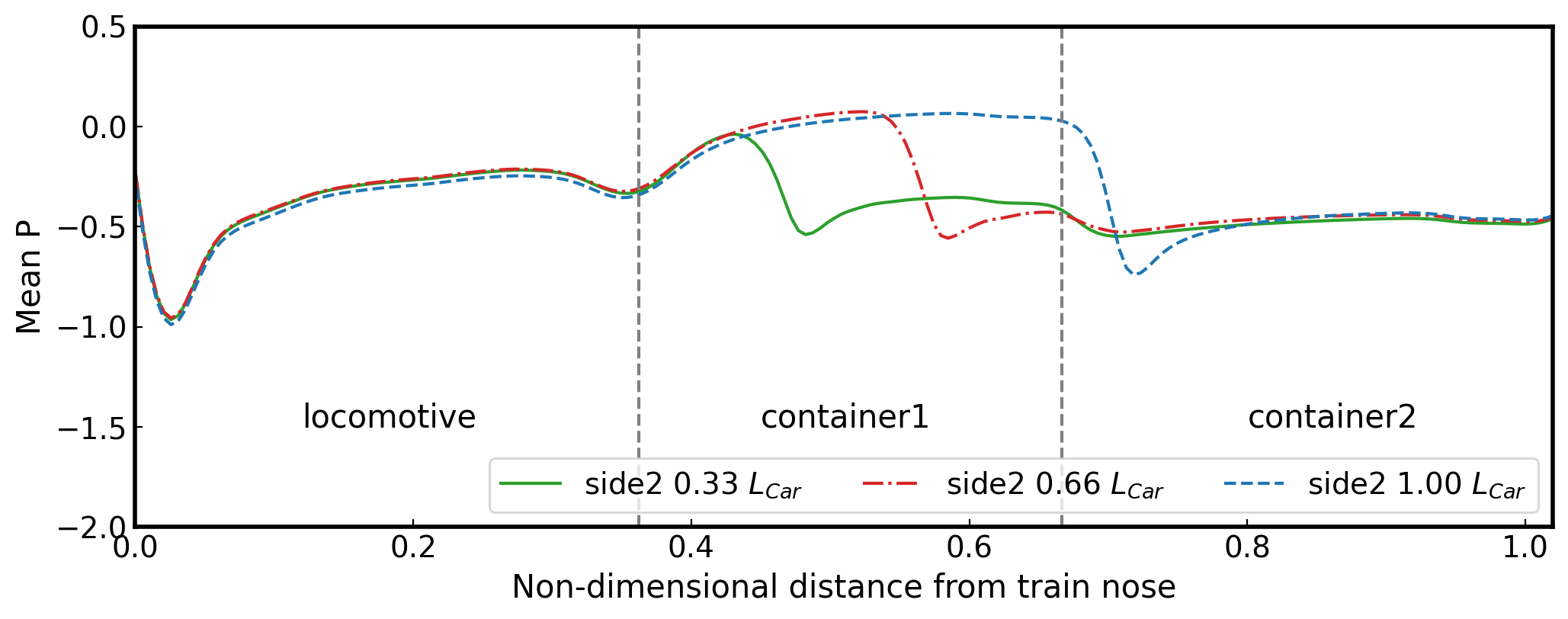}
    \caption*{\text{\small (d)}}
    \end{subfigure}
    \caption{Distribution of mean pressure at different lines positioned along the train (a) top1 (b) top2 (c) top3 (d) side2.}
    \label{fig:meanp_lines_streamwise}
\end{figure}

 To quantitatively compare the pressure variance, the mean pressure coefficient at different positions along the measuring lines is presented in Figure \ref{fig:meanp_lines_streamwise}. The positions of these measuring lines are illustrated in Figure \ref{fig:line_positions}, with the probes placed around the upper and lateral surfaces of the containers to capture variations due to flow separation and pressure recovery.

 As shown in Figure~\ref{fig:meanp_lines_streamwise}, the pressure profiles exhibit noticeable sensitivity to the gap length between containers. For instance, at position top1 and top2, a pronounced pressure drop is observed at the front of the second container in the case with an empty first wagon, indicating strong flow separation. As the gap reduces to 0.66 and 0.33 $L_\text{car}$, the adverse pressure gradient weakens and the pressure recovery along the top surface becomes smoother. Similar trends are observed at top3 and side2, confirming that increased gap length leads to stronger pressure disturbances and more extensive separation regions. These findings reinforce the observations made in Figure~\ref{fig:p_container1}, and further support the gap-dependent parameterisation of local aerodynamic effects for use in the 1D modelling framework.

\subsection{Parameterisation Equations} \label{chap:parameterisation_equation}
\begin{figure}[H]
    \centering
    \includegraphics[width=14cm]{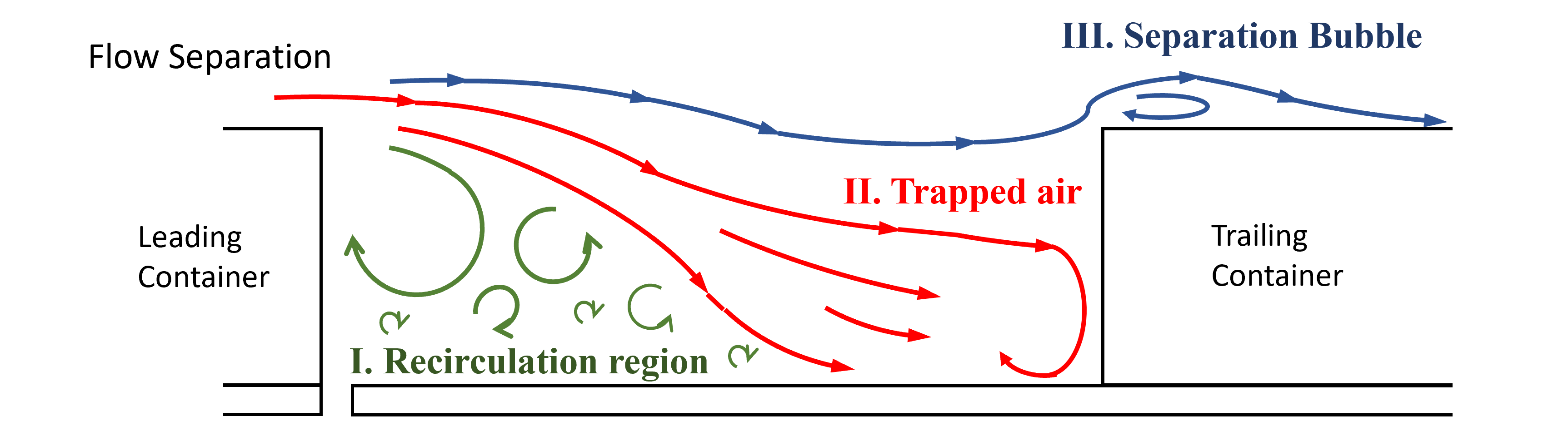}
    \caption{Diagram of flow characteristics inside the gap. }
    \label{fig:diagram_loading_configuration}
\end{figure}

Following the discussion above, the model of flow motion over and inside the gap between two containers can be characterised into three parts, and a diagram visualising these regions is illustrated in Figure~\ref{fig:diagram_loading_configuration}.

\begin{enumerate}[label=\Roman*]
\item The recirculation region dominated by large wake vortices generated from the separation of the boundary layer from the leading container.
\item The air flow downwash from the tail of the leading container, blocked by the trailing container, forming a recirculation in front of the trailing container and moving with the train as trapped air in the gap.
\item The outer part of the air can escape the gap and flow over the trailing container, forming a separation bubble region behind the sharp leading edge of the trailing container.
\end{enumerate}

It is important to highlight that both region I and II contribute to the increase in the effective cross-sectional area of the unloaded part of the train. In section~\ref{chap: parameterisationStudy_loadingconfiguration}, through flow visualisation and analysis, it was observed that for the case with a gap of 33 $\%$ $L_{\text{car}}$, the bottom of the trailing container is positioned at the transition between region I and II. Furthermore, for the case with a gap of 66 $\%$ $L_{\text{car}}$, the bottom of the trailing container is within region II. Consequently, the total effective blockage area for both cases is the sum of regions I and II. In section~\ref{Chap:Loading_configuration_modification}, the effective blockage areas for 33 $\%$ $L_{\text{car}}$ and 66 $\%$ $L_{\text{car}}$ were determined using the best fit method, resulting in values of $0.85 \times E_{\text{con}}$ and $0.7 \times E_{\text{con}}$ respectively. Additionally, for a fully loaded freight train (gap length is close to zero), the effective cross-sectional area equals $1 \times E_{\text{con}}$. Now, assuming the upper edge of the trapped air flow trajectory is approximately a straight line, then it is possible to approximate a linear relationship between the gap length and the effective blockage area. Thus one could estimate the fitting equation between gap length $L_{\text{gap}}$ and effective blockage area at the gap,
\begin{equation}
    E_{\text{gap}}=(1-0.4\frac{L_{\text{gap}}}{L_{\text{car}}})E_{\text{con}}
    \label{eq:gap_area}
\end{equation}

\begin{figure}[H]
    \centering
    \begin{subfigure}[t]{0.28\textwidth}
    \centering
    \includegraphics[width=\textwidth]{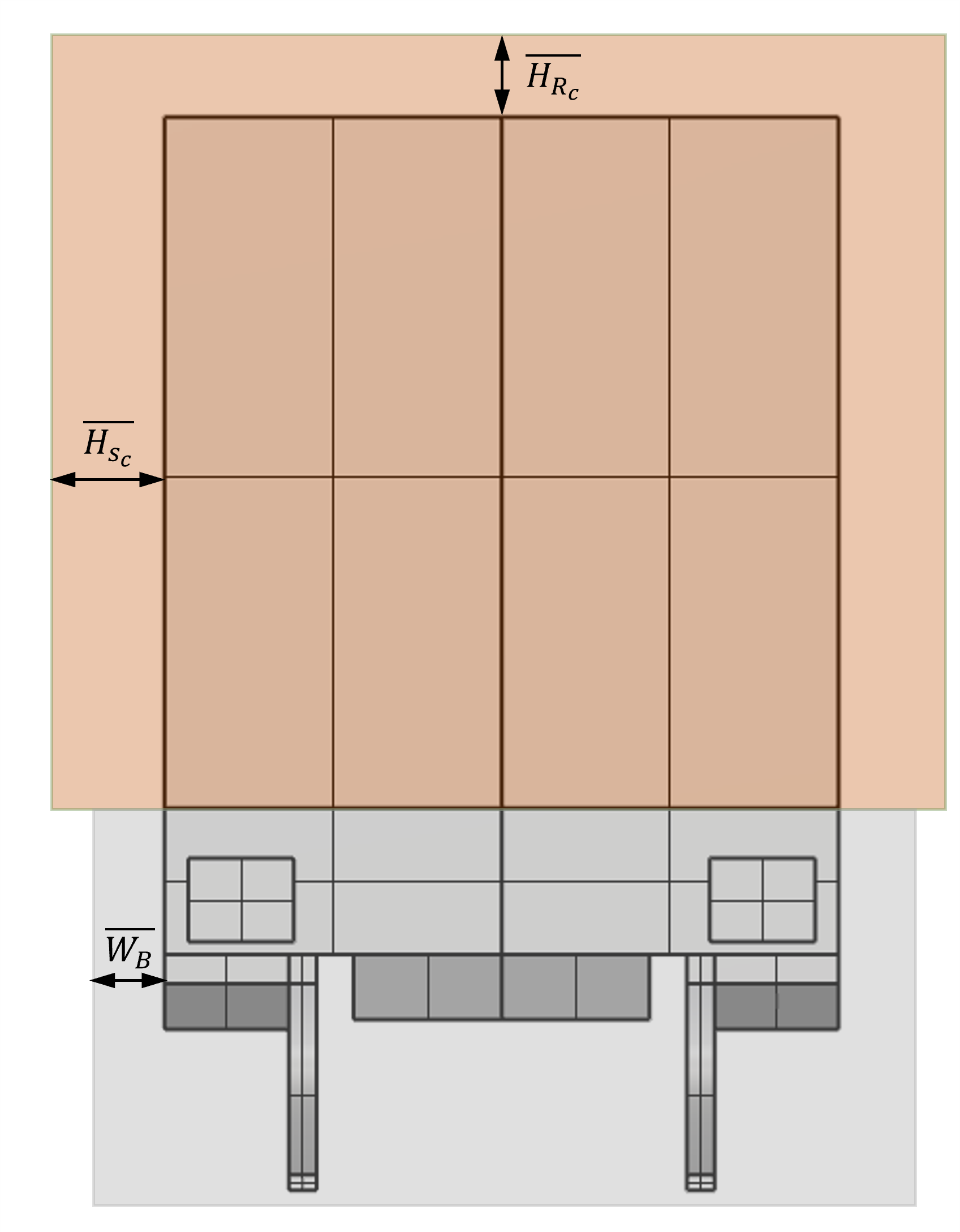}
    \caption*{\text{\small (a)}}
    \end{subfigure}
    \begin{subfigure}[t]{0.7\textwidth}
    \centering
    \includegraphics[width=\textwidth]{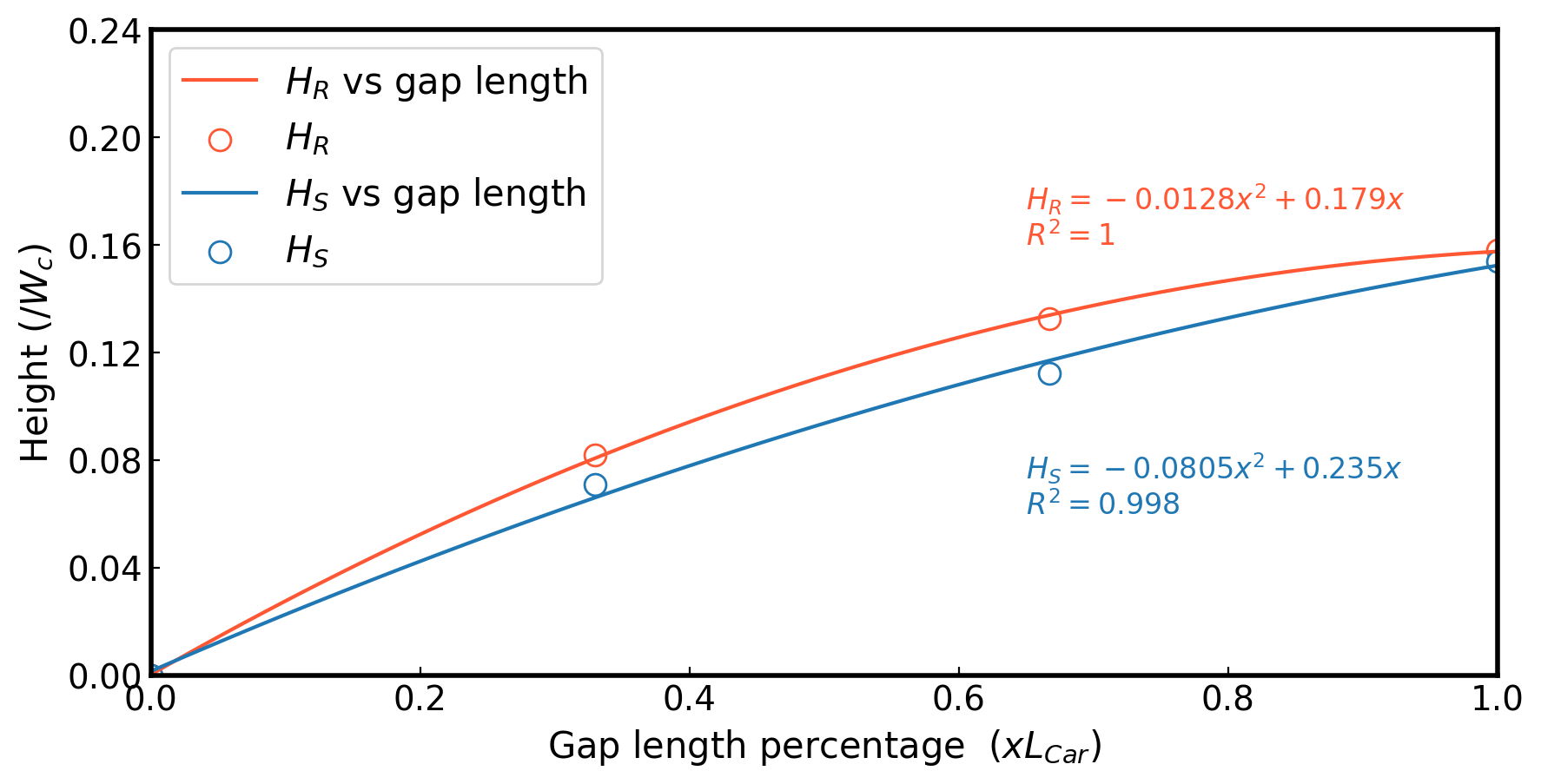}
    \caption*{\text{\small (b)}}
    \end{subfigure}
    \caption{Estimation of the loaded container's effective area.}
    \label{fig:seperation_container}
\end{figure}

Region III represents the separation bubble region around the trailing container, and its size significantly impacts the pressure wave rise when the container enters the tunnel, especially for sufficiently long gaps where the bubble size cannot be neglected. The separation bubble model for the containers is similar to that for the locomotive. The methodology to estimate the cross-sectional area of the loaded wagon considering the influence of separation bubble is illustrated in Figure~\ref{fig:seperation_container}(a). The effective cross-sectional area of the bogie is the same as that used for the locomotive as given in previous study by Liu et al. \cite{liu2023study}. The effective blockage area of the container part is estimated to be rectangular considering the separation bubble height $H_R$ and $H_S$. Using the data in section~\ref{chap: parameterisationStudy_loadingconfiguration}, the relationship equations for the different heights of separation bubbles at the top and side of the container and the percentage gap length are consolidated in Figure~\ref{fig:seperation_container}(b). Thus by utilising the separation bubble height at different positions, one can estimate the effective cross-sectional area of the container$E_{eff}$, and thus obtain the radius of the separation bubble at the trailing container, denoted as $r_{b_C}$($E_{eff}= \pi (r_C+r_{b_C})^2$, $r_C$ is the the radius of the container calculated directly from the geometry), after a certain distance of an unloaded gap. The squeezing effect induced by the tunnel cross-sectional area on $r_{b_C}$ can be considered analogous to that on $r_b$ (the effective radius of separation bubble at freight locomotive). Consequently, the scale factor for the separation bubble formed at the container $C_{b_C}$ effectively shares the same relationship equation for the locomotive (eq.~\ref{eq:blocakge_ratio}) \cite{liu2023study}, where $\beta_{E_C}$ is the blockage ratio at container.
\begin{equation}
    C_{b_C} = 1.75 \beta_{E_C} +1
    \label{eq:blocakge_ratio}
\end{equation}
\begin{figure}[htbp]
    \centering   
    \includegraphics[width=12cm]{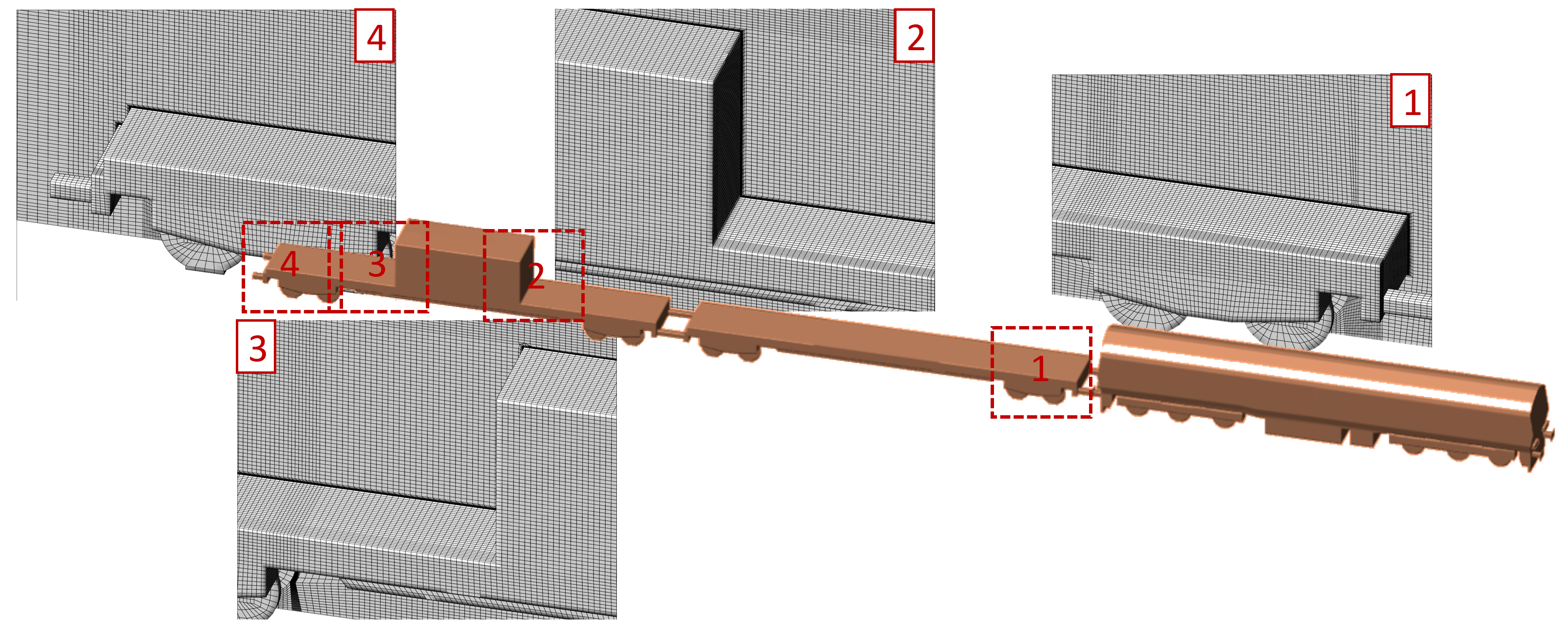}
    \caption{Surface mesh of freight train for validating loading configuration.}
    \label{fig:Loading_1.33L}
\end{figure}

\subsection{Validation of the Model} \label{chap:validation}
To evaluate the performance of the proposed methodology, the pressure time history curves obtained from the redeveloped 1D numerical model are compared against the CFD results using the LES model once again, but for a different train formation. A freight train with new loading configuration, in which the first carriage is fully unloaded and the second carriage 33$\%$ loaded, is adopted for this simulation. Furthermore, a new tunnel configuration is adopted, with a cross-sectional area of 70$m^2$. The mesh of the train surface and the wagon/container loading situation of the freight train are illustrated in Figure~\ref{fig:Loading_1.33L}.

The cross-sectional area, length and separation bubble $r_{b}$ of the Class 66 locomotive used in the 1D programme are the same as given in \cite{liu2023study}.The effective cross-sectional area of the gap between the locomotive and the first loaded container is calculated using the fitting Equation~\ref{eq:gap_area}, which is approximately $0.418\times E_{\text{con}}$. The separation bubble heights at the top and side of loaded container behind the gap are calculated using the parameterisation equation given in Figure~\ref{fig:seperation_container}. The reduced scale factor $C_{b_C}$ for the effective blockage area of the container inside a $70 m^2$ tunnel is calculated using Equation~\ref{eq:blocakge_ratio}, which is found to be 0.76 with ${\beta_{E_C}=0.136}$. Subsequently, $r_{b_C}$ can be obtained from the effective blockage area of the container.  

\begin{figure}[H]
    \centering   
    \begin{subfigure}[t]{0.49\textwidth}
    \centering
    \includegraphics[width=\textwidth]{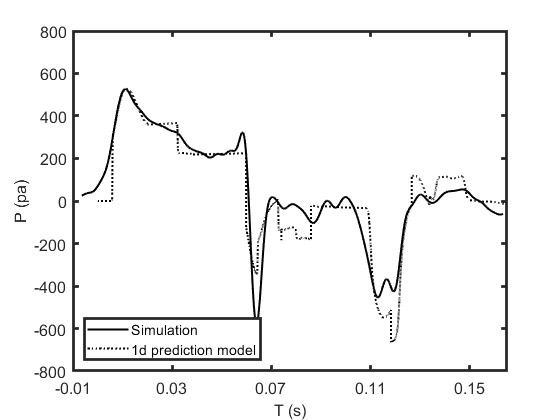}
    \caption*{\text{\small (a)}}
    \end{subfigure}
    \begin{subfigure}[t]{0.49\textwidth}
    \centering
    \includegraphics[width=\textwidth]{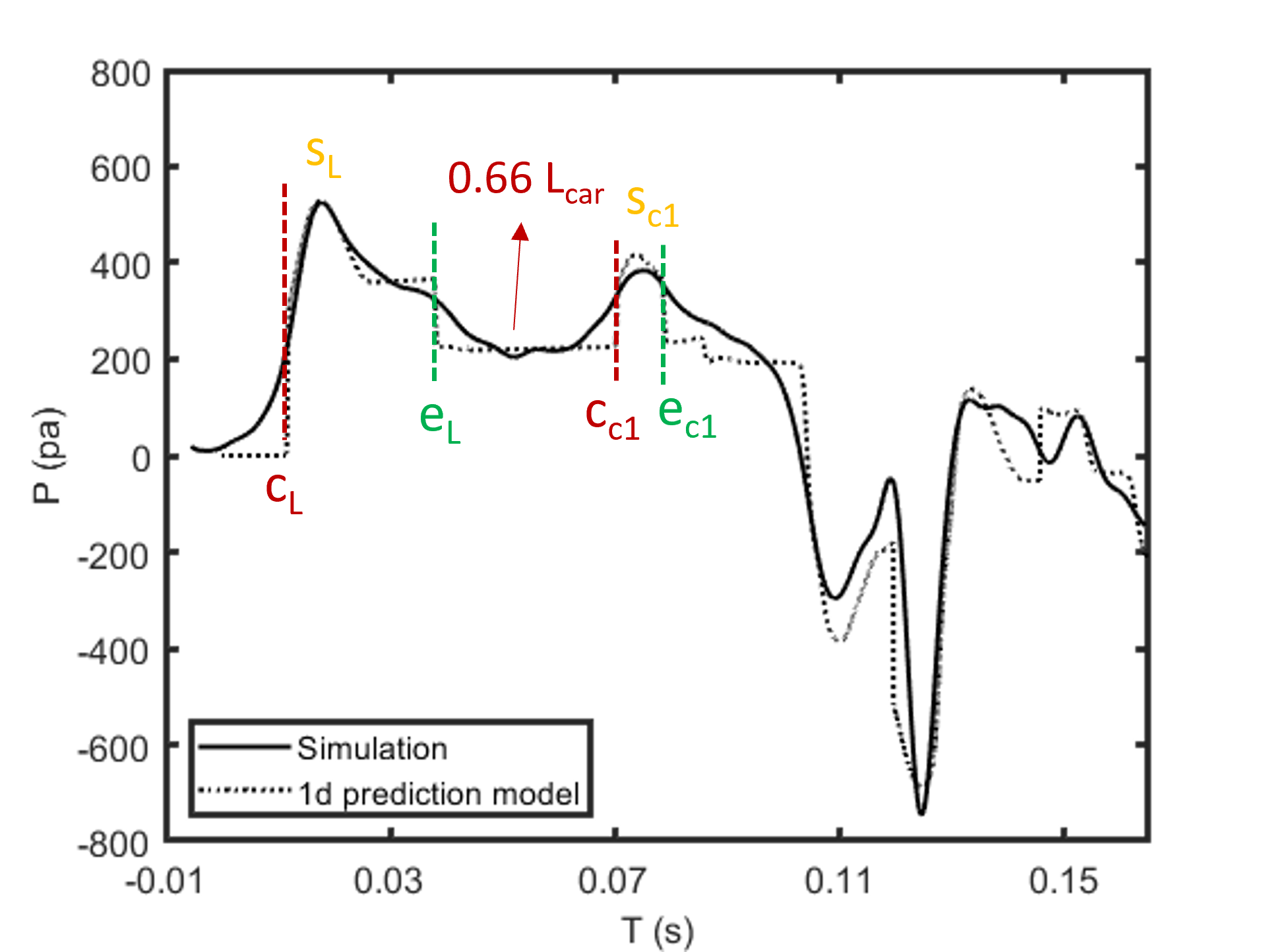}
    \caption*{\text{\small (b)}}
    \end{subfigure}
    \begin{subfigure}[t]{0.49\textwidth}
    \centering
    \includegraphics[width=\textwidth]{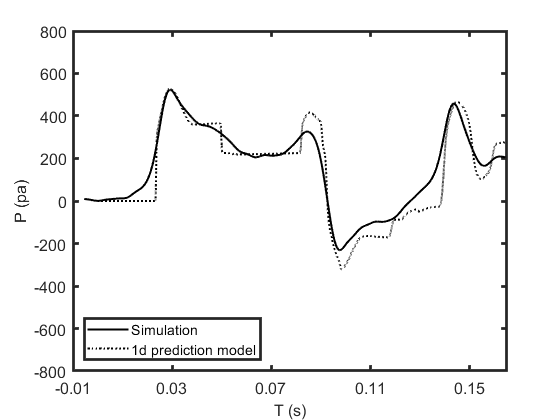}
    \caption*{\text{\small (c)}}
    \end{subfigure}
    \begin{subfigure}[t]{0.49\textwidth}
    \centering
    \includegraphics[width=\textwidth]{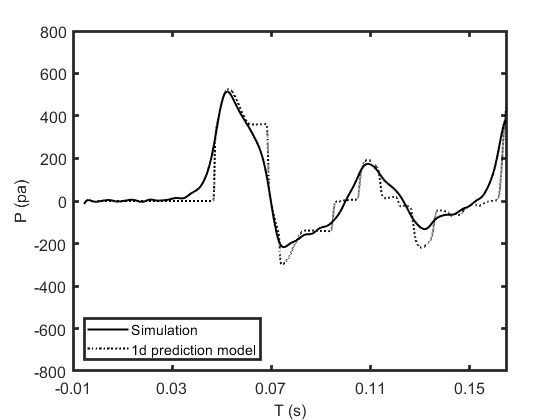}
    \caption*{\text{\small (d)}}
    \end{subfigure}
    \caption{Comparison of pressure history curves for the validation case predicted by the redeveloped 1D programme and LES simulation at (a) 2 m (b) 4 m (c) 8 m (d) 16 m from tunnel entrance.}
    \label{fig:Loading_configuration_validation2}
\end{figure}

Incorporating the predetermined parameters into the 1D programme, the comparison of pressure history curves at measuring points at 2 m, 4 m, 8 m, and 16 m inside the tunnel are illustrated in Figure~\ref{fig:Loading_configuration_validation2}. With the integration of these parameterisation models into the modified 1D programme, it can be seen that the programme can accurately predict the pressure wave patterns generated by freight trains under a certain range of loading configurations. Specifically, after accounting for the wake region generated behind the first locomotive and considering the effective cross-sectional area of the gap as approximately $0.418 \times E_{\text{con}}$, the pressure drop caused by the expansion wave when the tail of the locomotive enters the tunnel $e_{L}$, as calculated from the 1D code, is approximately 180 Pa, which aligns well with the LES results. Additionally, the pressure rise due to the compression wave $c_{c1}$ and the change in the pressure wave pattern resulting from the separation around the first container, $s_{c1}$, as shown in Figure~\ref{fig:Loading_configuration_validation2}(b), also exhibit strong agreement with the LES result. However, limitations exist in the broader application of this parameterisation model due to computational and experimental constraints. Nonetheless, this code serves as a foundational framework for further development of the model under various working conditions in future research.

\section{Conclusion}
\textcolor{black}{A thorough investigation was performed to simulate the pressure wave formation for a partially loaded intermodal container freight trains entering tunnels. The study introduced a redeveloped 1D model to accurately capture the effects of discontinuities caused by partial container loading, validated through LES simulations. Key contributions include:}

\begin{enumerate}[label=(\alph*)]

\item \textcolor{black}{The characteristic ‘step-like’ pressure rise is attributed to compression and expansion waves caused by abrupt area changes from the unloaded gaps. The flow in these gaps is classified into three regions: (i) wake recirculation behind the leading container, (ii) airflow down-wash, and (iii) outer airflow forming a separation bubble. These regions collectively affect the effective blockage and pressure wave strength.}

\item \textcolor{black}{The direction of wall shear stress in the wake region is the same as the direction of train operation. Separation regions form not only in the gaps but also ahead of blunt containers, reducing the effective frictional coefficient compared to fully loaded cases.}

\item \textcolor{black}{Development and validation of an improved 1D computational model incorporating a new mesh system and boundary conditions to accurately predict pressure variations for partially loaded configurations. Its predictions agree well with LES results for a 33\% loading case.}

\item \textcolor{black}{Quantitative analysis of flow regions, noting distinct velocity profiles caused by separation bubbles in different loading setups.}

\item \textcolor{black}{As the gap length increases, the effective blockage area decreases, while the airflow velocity in front of the trailing container approaches free stream conditions. This enhances separation bubble formation, whose height and evolution are quantitatively analysed.}

\item \textcolor{black}{A parameterisation study was performed to establish empirical relationships between key aerodynamic parameters and gap length. These relationships were embedded into the 1D model and successfully validated through additional simulation, allowing rapid estimation of pressure wave evolution under varied loading configurations.}

\end{enumerate}

\bibliographystyle{elsarticle-num-names}

\bibliography{references}

\end{document}